\documentclass[12pt,preprint]{aastex}

\begin{document}

\title{A \emph{Spitzer} MIPS Study of 2.5-2.0 $M_{\sun}$ Stars in Scorpius-Centaurus}
\author{Christine H.\ Chen\altaffilmark{1}, 
        Mark Pecaut\altaffilmark{2}, Eric E. Mamajek\altaffilmark{2,3},
        Kate Y. L. Su\altaffilmark{4}, Martin Bitner}
\altaffiltext{1}{Space Telescope Science Institute, 3700 San Martin Dr., Baltimore, MD 21218; cchen@stsci.edu}
\altaffiltext{2}{Department of Physics and Astronomy, University of Rochester, Rochester, NY 14627}
\altaffiltext{3}{Cerro Tololo Inter-American Observatory, Casilla 603, La Serena, Chile}
\altaffiltext{4}{Steward Observatory, University of Arizona, 933 North Cherry Avenue, Tucson, AZ 85721}                 

\begin{abstract}
We have obtained \emph{Spitzer Space Telescope} Multiband Imaging Photometer for \emph{Spitzer} (MIPS) 24 $\mu$m and 70 $\mu$m observations of 215 nearby, \emph{Hipparcos} B- and A-type common proper motion single and binary systems in the nearest OB association, Scorpius-Centaurus. Combining our MIPS observations with those of other ScoCen stars in the literature, we estimate 24 $\mu$m B+A-type disk fractions of 17/67 (25$^{+6}_{-5}$\%), 36/131 (27$^{+4}_{-4}$\%), and 23/95 (24$^{+5}_{-4}$\%) for Upper Scorpius ($\sim$11 Myr), Upper Centaurus Lupus ($\sim$15 Myr), and Lower Centaurus Crux ($\sim$17 Myr), respectively, somewhat smaller disk fractions than previously obtained for F- and G-type members. We confirm previous \emph{IRAS} excess detections and present new discoveries of 51 protoplanetary and debris disk systems, with fractional infrared luminosities ranging from $L_{IR}/L_{*}$ = 10$^{-6}$ to 10$^{-2}$ and grain temperatures ranging from $T_{gr}$ = 40 - 300 K. In addition, we confirm that the 24 $\mu$m and 70 $\mu$m excesses (or fractional infrared luminosities) around B+A type stars are smaller than those measured toward F+G type stars and hypothesize that the observed disk property dependence on stellar mass may be the result of a higher stellar companion fraction around B- and A-type stars at 10 - 200 AU and/or the presence of Jupiter-mass companions in the disks around F- and G-type stars. Finally, we note that the majority of the ScoCen 24 $\mu$m excess sources also possess 12 $\mu$m excess, indicating that Earth-like planets may be forming via collisions in the terrestrial planet zone at $\sim$10 - 100 Myr.
\end{abstract}


\section{Introduction}
Radial velocity studies of solar-like stars and "retired" intermediate-mass stars have discovered hundreds of Jovian-like planets and revealed correlations between host star properties and planet frequency. Measurements of [Fe/H] suggest that stars with super-solar metal abundance ($>$0.3 dex) are almost 10$\times$ more likely to possess gas giant companions than those with subsolar abundance (-0.5$<$[Fe/H]$<$0.0) \citep{fischer05}. Studies of host star mass suggest that the fraction of stars with Jovian planet companions increases as a function of stellar mass with intermediate mass stars (2 $M_{\sun}$) possessing twice as many companions on average compared to solar-like stars \citep{johnson10}. Since planets are believed to form in circumstellar disks around young stars, planetary demographics in conjunction with disk observations, are expected to place constraints on the processes by which planets form. We have conducted a search for dusty disks around young stars in ScoCen to determine whether disk properties (e.g. frequency, fractional infrared luminosity, grain temperature) are consistent with oligarchic growth models and dependent on stellar host properties.

The Scorpius-Centaurus OB association (ScoCen), with typical stellar distances of $\sim$100 - 200 pc, is the closest OB association to the Sun and contains three subgroups: Upper Scorpius (US), Upper Centaurus Lupus (UCL), and Lower Centaurus Crux (LCC), with estimated ages of $\sim$11 Myr \citep{pecaut12}, $\sim$15 Myr, and $\sim$17 Myr \citep{mamajek02}, respectively. The close proximity of ScoCen and the age of its constituent stars make this association an excellent laboratory for studying the formation and evolution of planetary systems. Several hundred candidate members have been identified to date; although, the association probably contains thousands of low-mass members. Member stars with spectral type F and earlier have been identified using moving group analysis of \emph{Hipparcos} positions, parallaxes, and proper motions \citep{dezeeuw99}, while later type members have been identified using youth indicators (i.e., high coronal X-ray activity and large lithium abundance; \cite{preibisch08,slesnick06}).

Infrared surveys of $\sim$10 - 20 Myr old moving groups and clusters indicate that young stars possess disks with a wide variety of properties. A \emph{Spitzer} IRAC and IRS peak-up survey of 204 B- through M-type (0.1 - 20 M$_{\sun}$) stars in Upper Sco found that the disks around late-type K- and M-type members are optically thick and accreting while  those around B- and A-type members are optically thin and gas-poor, consistent 
with the expectation that disks evolve faster around higher mass stars \citep{carpenter06}. High-resolution imaging studies of TW Hya and HR 4796A in the $\sim$10 Myr old TW Hya Association suggest that 10 - 20 Myr old disks may be sculpted by planets regardless of the mass of the central star or the evolutionary state of the disk. VLA thermal emission mapping has revealed the presence of a $\sim$4 AU inner hole in the TW Hya disk that may be dynamically cleared by a forming giant planet \citep{hughes07}. \emph{HST} STIS scattered light imaging has revealed the presence of a steep inner truncation at $\sim$65 AU in the ring-like debris disk around HR 4796A that is consistent with the presence of one or more planetary-mass companions \citep{schneider09}.

\cite{kenyon04,kenyon08} have developed "self-stirred" disk models to describe the formation of oligarchs and their impact on disk evolution around solar to intermediate mass stars with ages $>$few million years. They assume that disks are initially gas-rich and possess planetesimals with radii 1 m - 1 km at 30 - 150 AU from the central star. Since the disks are initially gas rich, the relative velocities between planetesimals are small, leading to constructive collisions between planetesimals that continue to grow until they have formed oligarchs (1000 km-sized bodies) that have depleted their feeding zone. The oligarchs are then massive enough to gravitationally perturb remnant planetesimals into eccentric orbits where they collide at high relative velocities, generating micron-size sized grains that may be detected via thermal emission. \cite{kenyon04,kenyon08} predict that the number of micron-sized debris particles and therefore the infrared excess emission rises rapidly at 5 Myr and peaks at 10-15 Myr. Their models assume that the total planetesimal mass is proportional to the stellar mass, suggesting that the disks around intermediate-mass stars are dustier and brighter than those around solar-like stars. Their models also indicate that evolutionary timescale for disks around intermediate-mass stars is faster than that around solar-like stars because the dynamical timescale around higher mass stars is shorter. 

Several groups have used \emph{Spitzer} MIPS at 24 $\mu$m to search for infrared excess trends expected from self-stirred disks around intermediate mass stars at 5 - 100 Myr. The first such study was carried out by \citet{hernandez06} who compared measurements of  26 early-type (F5 or earlier) stars in the $\sim$5 Myr Orion OB1a and 34 early-type stars in the $\sim$10 Myr Orion OB1b at $\sim$400 pc with observations of intermediate-mass stars in young clusters. Based on a modest sample of stars, they concluded that 24 $\mu$m excess peaks at 10 Myr, consistent with the onset of oligarchic growth at 30-150 AU. \citet{currie08} carried out a larger survey of $\sim$600 intermediate and high-mass stars in h and $\chi$ Persei at $\sim$2.5 kpc. Their statistical analysis also indicated a peak in 24 $\mu$m excess emission at 10-15 Myr, followed by a decline up to 1 Gyr. More recently, \citet{carpenter09} observed 62 B- and A-type members of the $\sim$11 Myr Upper Sco. They found that an increase in 24 $\mu$m excess emission for stars with ages between 5 and 17 Myr was statistically insignificant ($<$2$\sigma$).

We report the results of a \textit{Spitzer} MIPS 24 $\mu$m and 70 $\mu$m survey of 215 B- and A-type \emph{Hipparcos} common proper motion members of ScoCen stars, expanding on our survey of 182 F- and G-type members \citep{chen11}. We list the targets for the full sample, along with their spectral types, distances, and subgroup memberships in Table \ref{tab:starprops}. We compare our data with (1) \cite{kenyon04,kenyon08} self-stirred disk models to determine whether models of 10-20 Myr disks are consistent with our data and (2) MIPS 24 $\mu$m and 70 $\mu$m photometry of F- and G-type members \citep{chen11} to search for evidence of stellar mass dependent disk evolution, expected based on results from radial velocity surveys. A recent cursory analysis of the preliminary WISE ScoCen data indicates that the disks around intermediate type members are less massive than those around solar-like members \citep{rizzuto12}.

\section{Observations}
The \emph{Hipparcos} satellite enabled high-precision measurements of stellar position, parallax, and proper motion for stars with V-band magnitudes, $m_{V}$ $<$ 9, providing the ability to efficiently identify candidate members of OB associations kinematically with spectral types later than B for the first time. \cite{dezeeuw99} (hereafter dZ99) analyzed the \emph{Hipparcos} measurements of OB associations using the \cite{debruijne99} refurbished convergent point method and the \cite{hoogerwerf99} "Spaghetti method" to determine the average position and space motions of each association, including the US, UCL, and LCC subgroups of ScoCen. By cross correlating the \emph{Hipparcos} measurements of individual stars with mean subgroup properties, dZ99 carried out a detailed census of high- and intermediate-mass stars in ScoCen. In particular, they identified more than 300 probable new intermediate-mass members of US (49 B-type and 34 A-type stars), UCL (66 B-type and 68 A-type stars), and LCC (42 B-type and 55 A-type stars). However, they cautioned that up to $\sim$30\% of their candidate members may be interlopers because the stellar radial velocities were not measured. Recently, \cite{rizzuto11} re-examined the membership of high mass stars in the ScoCen region (285$\arcdeg$ $\leq$ $l$ $\leq$ 360$\arcdeg$, -10$\arcdeg$ $\leq$ $b$ $\leq$ 60$\arcdeg$), estimating membership probabilities based on stellar distances, Galactic velocities, latitudes, and longitudes inferred from \emph{Hipparcos} positions, proper motions, and parallaxes and 2nd Catalog of Radial Velocities with Astrometric Data (CRVAS-2) radial velocities, where available.

Since our survey was defined prior to the \cite{rizzuto11} analysis, we sought to obtain MIPS 24 $\mu$m and 70 $\mu$m photometry of all of the intermediate-mass ScoCen members identified by dZ99. Prior to our study, several groups had already begun to explore US in detail. \cite{oudmaijer92} had discovered \emph{IRAS} excesses associated with 2 B-type (HIP 80569/HD 148184 and HIP 81474/HD 149914) and 2 A-type (HIP 79476/HD 145718 and HIP 81624/HD 150193) binary and single star systems  in US. Using \emph{Spitzer} MIPS, \cite{carpenter09} had observed and 37 B-type and 25 A-type binary and single star systems in US using MIPS and \cite{su06} had observed 3 B-type stars in US and 10 B-type and 6 A-type stars in UCL. We have obtained \emph{Spitzer} MIPS observations of 1 A-type star in US, and 56 B-type and 62 A-type stars in UCL, completing the dZ99 sample of B- and A-type stars in UCL. We have also obtained observations of 42 B-type and 54 A-type members of LCC, all dZ99 B- and A-type members of LCC except one A-type star (HIP 78541/HD 143488). 

\subsection{MIPS Observations}
We obtained \textit{Spitzer} \citep{werner04} MIPS \citep{rieke04} observations of 215 candidate ScoCen single and binary systems in photometry mode at 24 $\mu$m and 70 $\mu$m (default scale). Each system was observed once between 2007 August  and 2008 September 2008, using 1 cycle of 3 s integrations at 24 $\mu$m and 1 - 6 cycle(s) of 10 s integrations at 70 $\mu$m, corresponding to on-source integration times of 24.1 s and 125.8 s - 754.8 s at 24 $\mu$m and 70 $\mu$m respectively. All the data were processed using the MIPS instrument team Data Analysis Tool \citep{gordon05} for basic reduction (dark subtraction, flat-fielding/illumination correction). A series of additional steps designed to provide homogeneous reduction for MIPS data was applied as part of a \emph{Spitzer} legacy catalog \citep{su10}. In short, a second flat field constructed from the 24 $\mu$m data itself was applied to all of the 24 $\mu$m data to remove scattered-light gradient and dark latency to improve sensitivity (e.g., \cite{engelbracht07}) except for observations that possess complex background. The known transient behaviors associated with the MIPS 70 $\mu$m array were removed by masking out bright sources in the field of view and time filtering the data (for details see \cite{gordon07}). The processed data were then combined using the World Coordinate System information to produce final mosaics with pixels half the size of the physical pixel scale. For 70 $\mu$m data, an additional outlier rejection was performed using the spatial redundancy of each processed data frame to further remove hot pixels in the data. This extra process can improve the data quality, especially for observations where sources are not detected (K. Y. L. Su et al., in preparation).

Since the majority of the sources in the sample are not resolved, we extract the photometry using point-spread function (PSF) fitting. The input PSF's were constructed using observed calibration stars and smoothed STinyTim model PSFs, and have been tested to ensure that photometric results are consistent with the MIPS calibration \citep{engelbracht07, gordon07}. The systematic errors were estimated based on the pixel-to-pixel variation on the source-free (PSF-subtracted) images. We also performed aperture photometry (using the multiple aperture setting in \cite{su06}. The aperture photometry measurements were used as a reference to screen targets that might be contaminated by nearby sources, background nebulosity, or source extension. We list our measurements in Table \ref{tab:mips}; stars with 24 $\mu$m contamination are noted with a dagger. All of the targets were detected at 24 $\mu$m. Accurate 24 $\mu$m photometry could not be measured for the heavily saturated Herbig Ae/Be star HIP 56379 (HD 100546).Each target position was refined using two-dimensional Gaussian fitting and then compared to the SIMBAD stellar position to ensure the correct source extraction. For the sources that were not detected at 70 $\mu$m, the PSF was fixed at the position of the 24 $\mu$m source position to extract PSF fitting photometry using the minimum $\chi^{2}$ technique. We quote 3$\sigma$ upper limits for systems that were not detected. The total photometric uncertainty is the sum in quadrature of (1) the source photon counting uncertainty, (2) the detector repeatability uncertainty (0.4\% and 4.5\% of the total flux at 24 and 70 $\mu$m, respectively), and (3) the absolute calibration uncertainty (2\% and 5\% of the total flux at 24 and 70 $\mu$m, respectively).

\section{Disk Fractions}

We estimate the stellar photospheric fluxes for our sample based on 2MASS \citep{cutri03} K$_{s}$-band magnitudes and intrinsic main sequence colors calculated by E. Mamajek\footnote{http://www.pas.rochester.edu/$\sim$emamajek/EEM$\_$dwarf$\_$UBVIJHK$\_$colors$\_$Teff.html}. First, we assembled the \emph{Hipparcos} B- and V-band photometry, Cousins I-band photometry (where available), and 2MASS J-, H-, and K$_{s}$-band photometry and constructed measured $B-V$, $V-I_{c}$, $V-J$, $V-H$, and $V-K_{s}$ colors. Second, we calculated the extinction for each star. For B-type stars with Stromgren photometry \citep{hauck98}, we estimated the extinction using the prescription of \cite{shobbrook83}; for B-type stars without Stromgren photometry but with UBV photometry \citep{mermilliod94}, we estimated the extinction using the reddening-free Q parameter, Q = (U-B) - 0.72*(B-V). For all remaining B-type stars and all the A-type stars, we calculated the extinction in each color assuming that the stars possess intrinsic main sequence colors, an average visual extinction, $A_{V}$, and its uncertainty based on the standard deviation of the extinction measurements. Fourth, we extrapolated the average $A_{V}$ to the $K_{s}$-band assuming that $A_{K}$ = 0.116 A$_{V}$, consistent with $R_{V}$ = 3.1 and a Cardelli, Clayton, \& Mathis (1989) extinction law. Fifth, we corrected the measured 2MASS $K_{s}$-band magnitudes for extinction. Finally, we used Kurucz models to estimate the predicted 24 $\mu$m and 70 $\mu$m fluxes based on the extinction-corrected K$_{s}$ band magnitudes and stellar spectral types, assuming the Kenyon \& Hartmann (1995) conversion between spectral type and effective temperature. For example, a star with $T_{eff}$ = 10,000 K, is expected to possess $F_{\nu}$($K_{s}$)/$F_{\nu}$(24 $\mu$m) = 96.7 and $F_{\nu}$($K_{s}$)/$F_{\nu}$(70 $\mu$m) = 890.0.

\begin{figure}
\figurenum{1}
\plotone{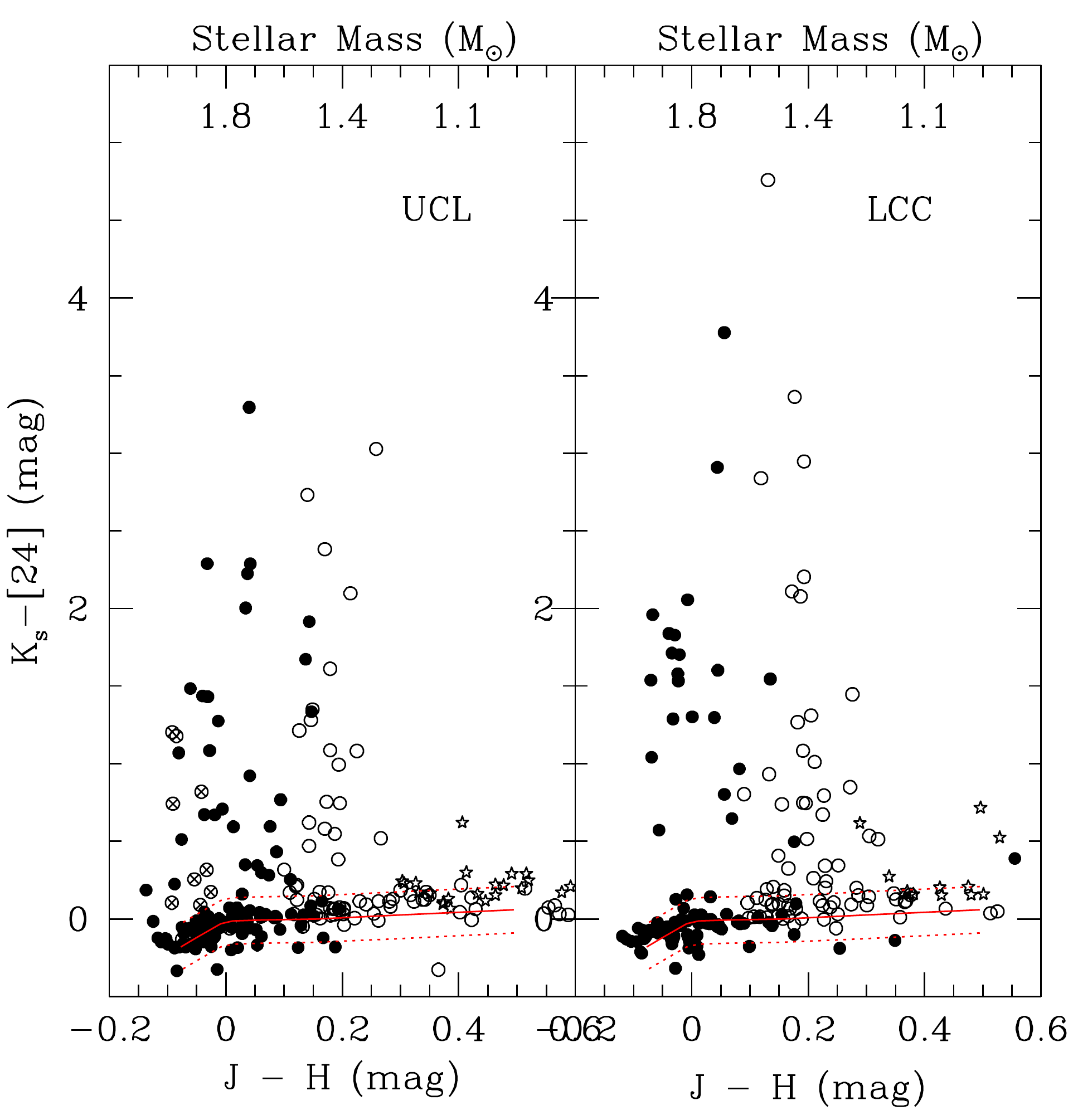}
\figcaption{The K$_{s}$ - [24] color plotted as a function of the J - H color for the Upper Centaurus Lupus (UCL) and Lower Centaurus Crux (LCC) subgroups of Sco Cen. Our sample of 209 B- and A-type stars is shown as filled circles; open circles with crosses represent  A-type stars from \citet{su06}; F- and G-type stars from \citet{chen11} are shown as open circles; and the \citet{carpenter08} sample of F- and G-type stars is shown as open stars. The solid line represents the colors for main sequence stars based on Kurucz models with log $g$ = 4.0 and solar metallicity. The dashed lines show the 3$\sigma$ range in K$_{s}$ - [24] color (0.15 mag) around the main sequence.
\label{fig:k24}}
\end{figure}

\begin{figure}
\figurenum{2}
\plotone{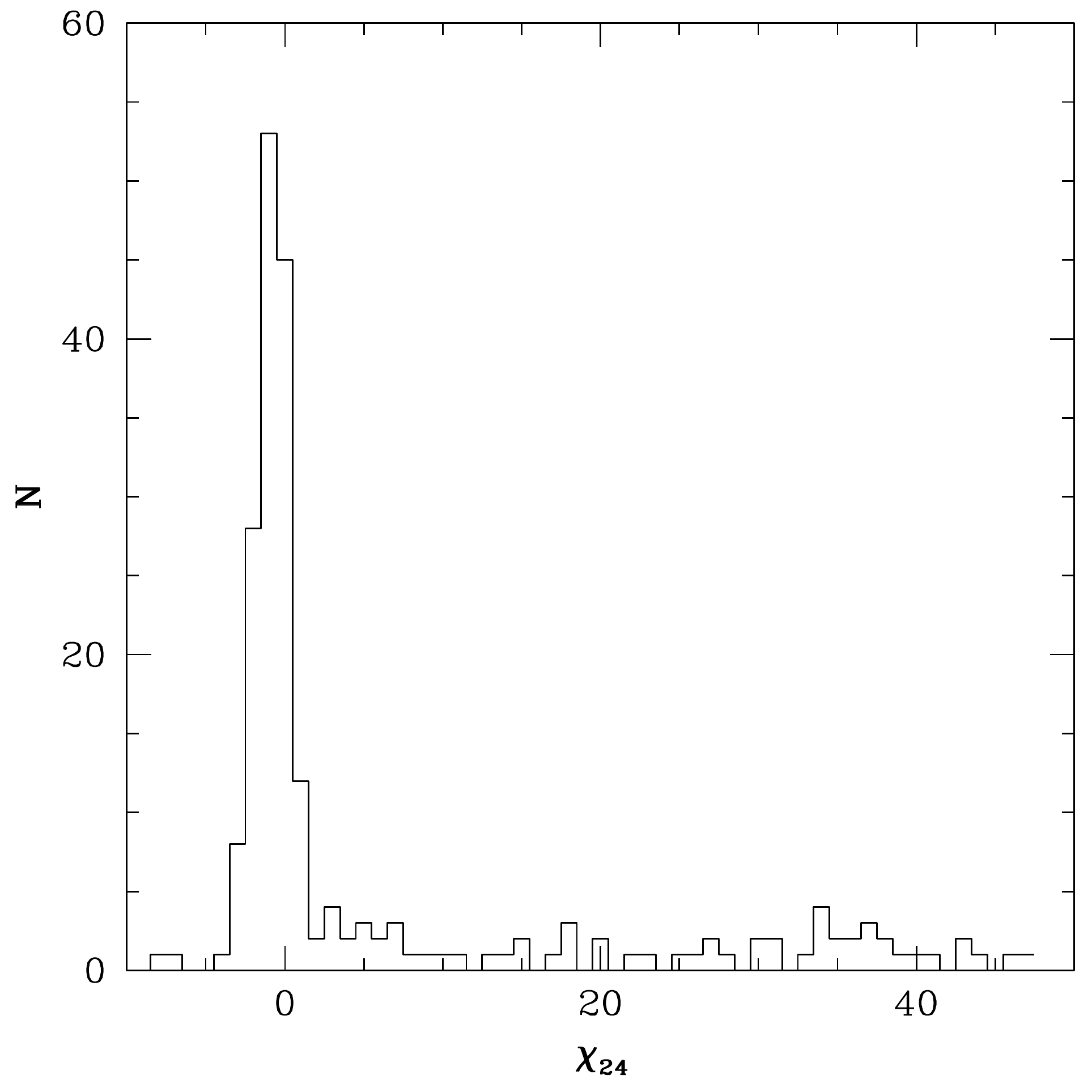}
\figcaption{Histogram of the significances ($\chi$ = ($F_{\nu}$(24 $\mu$m)-F$_{*}$(24 $\mu$m))/$\sigma_{F24}$) of observed 24 $\mu$m excesses. The histogram includes all sources in our study regardless of the membership in ScoCen. Sources with $\chi$ $\geq$ 6 also possess $K_{s}$ - [24] $>$ 0.30 mag nd were identified as having significant excess. 
\label{fig:chi24}}
\end{figure}

For comparison with our measured (but not color-corrected) fluxes, we list the predicted photospheric 24 and 70 $\mu$m fluxes integrated over the MIPS bandpasses, $F_{\nu}$, in Table \ref{tab:mips}. We calculate the excess significance of each detected source, $\chi$ = (Measured Flux - Predicted Flux)/Uncertainty, where the Uncertainty includes absolute calibration uncertainty (2\%), repeatability uncertainty (0.4\%), the photospheric model uncertainty ($\sim$3\%), and statistical uncertainty of the 24 $\mu$m photometry summed in quadrature. We list excess objects with $\chi$ $\geq$ 6 at 24 and/or 70 $\mu$m in Table \ref{tab:BBmodel}. We verify our stellar atmosphere model fits by examining the K$_{s}$ - [24] colors of our sources. All of our excess sources possess K$_{s}$ - [24] $>$ 0.2 mag. We plot the K$_{s}$ - [24] color as a function of J - H color (as a proxy for spectral type) for all of the sources in our study in Figure \ref{fig:k24}. We show the distribution of significance of the 24 $\mu$m excesses ($\chi_{24}$ = (Measured F$_{\nu}$(24 $\mu$m)- Predicted F$_{\nu}$(24 $\mu$m))/Measured $\sigma_{F24}$) in Figure \ref{fig:chi24}. Our K$_{s}$ - [24] selection criteria robustly identified sources with large excesses but may not identify sources with weaker excesses. Follow-up IRS spectroscopy for three of our excess sources indicates that they possess hydrogen emission lines, consistent with Be stars (HIP 63005, HIP 67472, and HIP 69618); the infrared excess associated with these disks is generated by excretion disks rather than debris disks.

\begin{figure}
\figurenum{3}
\plotone{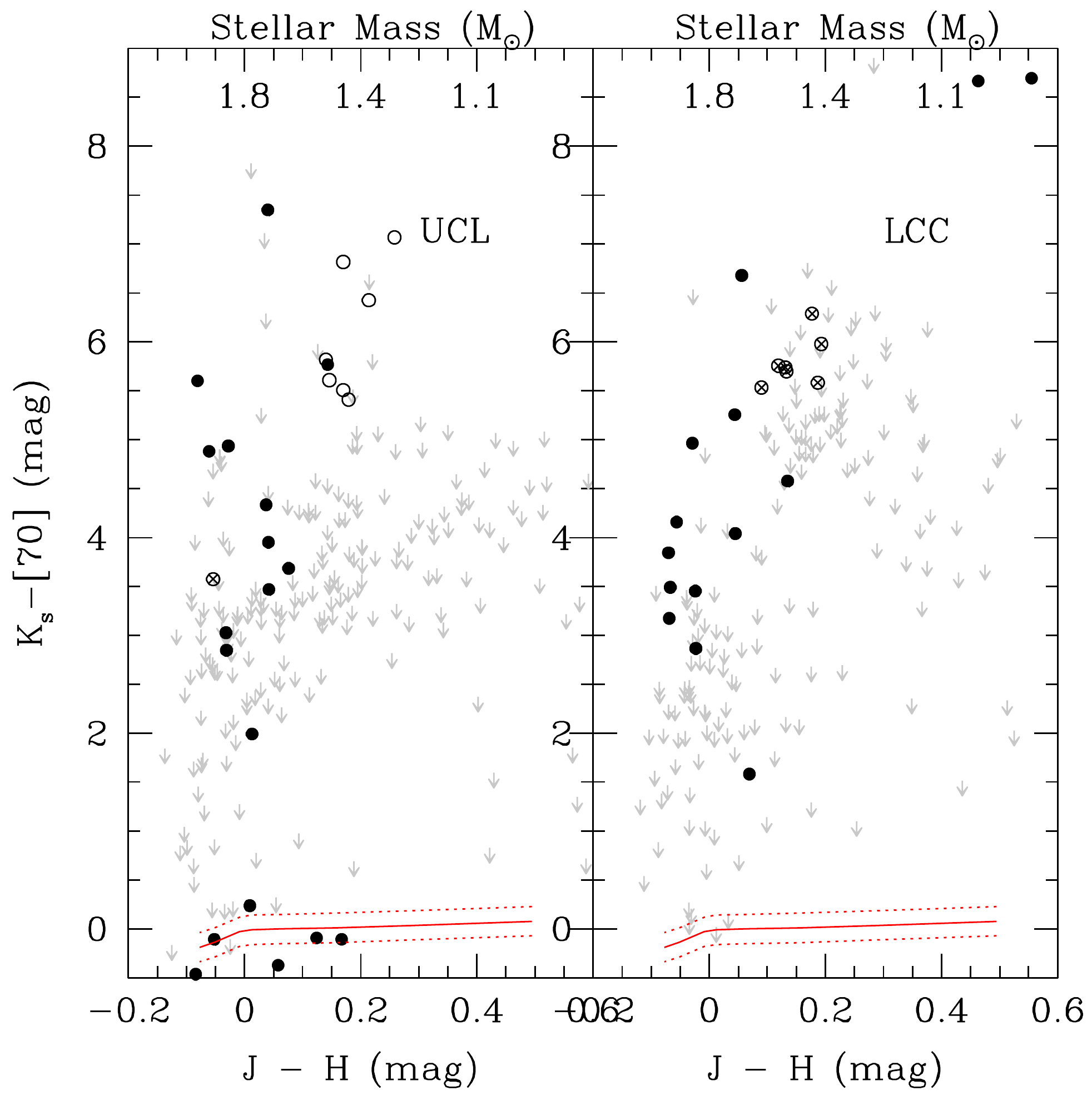}
\figcaption{K$_{s}$ - [70] color plotted as a function of J - H color for subgroups of Sco Cen.  The symbols are as described in Figure~\ref{fig:k24}. Non-detections are shown as gray upper limit symbols. The solid line represents the colors for main sequence stars based on Kurucz models with log $g$ = 4.0 and solar metallicity. The dashed lines show the 3$\sigma$ range in K$_{s}$ - [70] color (0.23 mag) around the main sequence.
\label{fig:k70}}
\end{figure}

Thirty-four sources, including twenty-seven 24 $\mu$m excess sources (not excluding Be stars), are detected at 70 $\mu$m. Since our 70 $\mu$m integration times were short, the majority of the objects detected at 70 $\mu$m possess strong 70 $\mu$m excesses; however, a handful of photospheres around bright B-type stars are also detected. We plot the K$_{s}$ - [70] color as a function of $J - H$ color for all of the sources in our study in Figure \ref{fig:k70}. For each 24 $\mu$m plus 70 $\mu$m excess source, we fit the MIPS 24 $\mu$m and 70 $\mu$m excess fluxes with a single-temperature blackbody, $T_{gr}$ (see Table \ref{tab:BBmodel}), and infer color temperatures $T_{gr}$ = 40 - 300 K and fractional infrared luminosities $L_{IR}/L_{*}$ = 1$\times$10$^{-5}$ to 4$\times$10$^{-2}$.  For each 24 $\mu$m excess only source, we cannot constrain the color temperatures without additional infrared excess detections at other wavelengths; however, we estimate grain temperature lower limits. Our 70 $\mu$m flux upper limits suggest that the color temperatures for these sources are consistent with those typically measured toward debris disks. For these sources, we infer infrared dust luminosities assuming that $F_{IR}$ $\sim$ $\nu F_{\nu}$(24 $\mu$m).

We measure B+A type 24 $\mu$m disk fractions for the two older ScoCen subgroups using data from our sample, excluding 3 Be stars in UCL and 1 Be star in LCC: 27/115 (23$^{+4}_{-3}$\%) for UCL ($\sim$15 Myr) and 23/95 (24$^{+5}_{-4}$\%) for LCC ($\sim$17 Myr), respectively. Since our census of LCC B+A stars is complete (except for one unobserved star), our estimated LCC B+A disk fraction accurately reflects the known demographics of the subgroup. We combine our dZ99 UCL B+A sample with that of \cite{su06} to determine a complete dZ99 UCL B+A disk fraction. We discovered 24 $\mu$m excess around 8/53 B- and 19/62 A-type dZ99 UCL stars, excluding Be stars; \cite{su06} discovered 24 $\mu$m excesses around 6/10 B- and 3/6 A-type dZ99 UCL stars, suggesting a combined dZ99 UCL B+A 24 $\mu$m excess fraction: 36/131 (27$^{+4}_{-4}$\%). Finally, we combine the \cite{oudmaijer92}, \cite{carpenter09}, and \cite{su06} dZ99 US B+A samples to determine the most complete dZ99 US B+A disk fraction. \cite{oudmaijer92} discovered \emph{IRAS} 25 $\mu$m excesses around 1/1 B-type and 2/2 A-type dZ99 US stars, excluding 1 Be star (HIP 80569); \cite{carpenter09} discovered MIPS 24 $\mu$m excesses around 6/35 B- and 7/25 A-type dZ99 US stars, excluding 2 Be stars (HIP 77859 and HIP 78207); \cite{su06} discovered MIPS 24 $\mu$m excesses around 0/3 B-type dZ99 US stars; we measured a MIPS 24 $\mu$m excess around 1 A-type dZ99 US star, suggesting a combined dZ99 US B+A 24 $\mu$m excess fraction: 17/67 (25$^{+6}_{-5}$\%). Since the dZ99 sample was selected on the basis of \emph{Hipparcos} position, parallax, and common proper motion, it is expected to be contaminated by interlopers. dZ99 estimate that (1-5)/49 B-type and  (2-6)/34 A-type stars in US, (5-12)/66 B-type and (10-15)/68 A-type stars in UCL, and (4-7)/42 B-type and (8-11)/55 A-type stars in LCC are interlopers, respectively. 

The evolution of disks around stars with masses 1-3 $M_{\sun}$ has been modeled by \cite{kenyon04,kenyon08}.  Approximately half (120/215) of the B- and A-type ScoCen members in our sample correspond to stars in this mass range (2.0-2.5 $M_{\sun}$). We searched for trends in the disk fraction around 2.0-2.5 $M_{\sun}$ as a function of time. By comparing the infrared excess emission of stars with different ages, we can quantify disk evolution. We expect that 2.5 $M_{\sun}$ stars will appear as B9.5 through B9 members of US, UCL, and LCC and that 2.0 $M_{\sun}$ stars will appear as A9 through A0 members of US and as A7 through A0 members of UCL and LCC (Table \ref{tab:clusters}). For US, we infer a disk fraction of 6/16 (38$^{+13}_{-10}$\%) for 2.5 $M_{\sun}$ stars and 7/20 (35$^{+12}_{-9}$\%) for 2.0 $M_{\sun}$ stars from the \cite{oudmaijer92} \emph{IRAS} 25 $\mu$m survey and the \cite{carpenter09} MIPS 24 $\mu$m survey. We note that 2 of the excess systems around A-type stars, discovered by \cite{oudmaijer92}, are "primordial" based on the presence of H$\alpha$ emission: HIP 79476/HD 145718 and HIP 81624/HD 150193 \citep{vieira03}. We measured 2.5 $M_{\sun}$ disk fractions of 6/20 (30\%$^{+12}_{-8}$\%) and 4/10 (40\%$^{+16}_{-12}$\%) in UCL and LCC from this survey and the \cite{oudmaijer92} \emph{IRAS} 25 $\mu$m survey, respectively, including one accreting, "primordial disk" in LCC (HIP 56379/HD 100546). We measured 2.0 $M_{\sun}$ disk fractions of 16/50 (32$^{+7}_{-6}$\%) and 15/49 (31\%$^{+7}_{-6}$\%) in UCL and LCC, respectively. Since the US disk fractions are similar to the UCL and LCC disk fractions (both including and excluding primordial systems), we do not believe that there is strong evidence for a change in the disk fractions between ages $\sim$11 Myr and $\sim$15-20 Myr for 2.0-2.5 $M_{\sun}$ stars. 

The MIPS 24 $\mu$m observations in our survey can be combined with those from other surveys \citep{chen11,carpenter09,su06,oudmaijer92} to determine whether ScoCen disks exhibit mass dependent evolution. We find that the disk fractions for 2.0-2.5 $M_{\sun}$ stars are somewhat smaller than those measured for 1.0-1.5 $M_{\sun}$ stars ($\sim$33\%) and that the disk fraction for higher mass stars ($>$3 $M_{\sun}$) is substantially smaller, 4/57 (7\%$^{+5}_{-2}$\%) with Be stars forming the majority of high mass systems with 24 $\mu$m and/or 70 $\mu$m excess. We plot the ratio of the observed and expected photospheric 24 $\mu$m and 70 $\mu$m fluxes for all of the stars observed in ScoCen as a function of stellar mass in Figure \ref{fig:r24} and Figure \ref{fig:r70}. Only HIP 81972 of the four high mass stars ($>$3 $M_{\sun}$) with 24 $\mu$m excess is not a Be star. High-resolution, ground-based, near-infrared imaging suggests that HIP 81972 is a multiple system with at least one confirmed companion with an estimated mass 0.35 $M_{\sun}$, located at an angular separation of 5.0$\arcsec$ \citep{kouwenhoven07}. Since the angular resolution of \emph{Spitzer} at 24 $\mu$m is $\sim$6$\arcsec$, we can not determine whether the primary star, HIP 81972A, or its confirmed late-type companion is the source of the detected infrared excess.

\section{Disk Evolution}
\cite{kenyon04,kenyon08,kenyon10} model the infrared excess emission from self-stirred disks around stars with masses 1-3 M$_{\odot}$ as a function of time, assuming that the disk is initially composed of planetesimals ranging in size from $\sim$1-1000 m at 30 - 150 AU. In these simulations, 1000 km planetary embryos first form at small radii where the surface density of solids is highest and then at subsequently greater distances with time. Once planetary embryos form, they trigger collisional cascades between nearby leftover planetesimals that grind remnant planetesimals into micron-sized dust grains that can be detected via thermal emission. Since the \cite{kenyon04,kenyon08,kenyon10} models initially include no primordial, micron-sized grains, they initially predict that the disks possess no infrared excess. As planetary embryo formation sweeps through the disk, the 24 $\mu$m flux ratio, $F_{\nu}$(observed)/$F_{\nu}$(predicted), increases until it reaches a peak around 5-20 Myr and then decreases as planetary embryo formation propagates into the outer regions of the disk where dust debris possesses thermal equilibrium temperatures, $T_{gr}$ $<$ 40 K, too cold to emit strongly at 24 $\mu$m. Since 70 $\mu$m excess emission is diagnostic of cooler dust, the models predict that the 70 $\mu$m excess declines more slowly with time.

\begin{figure}
\figurenum{6}
\plotone{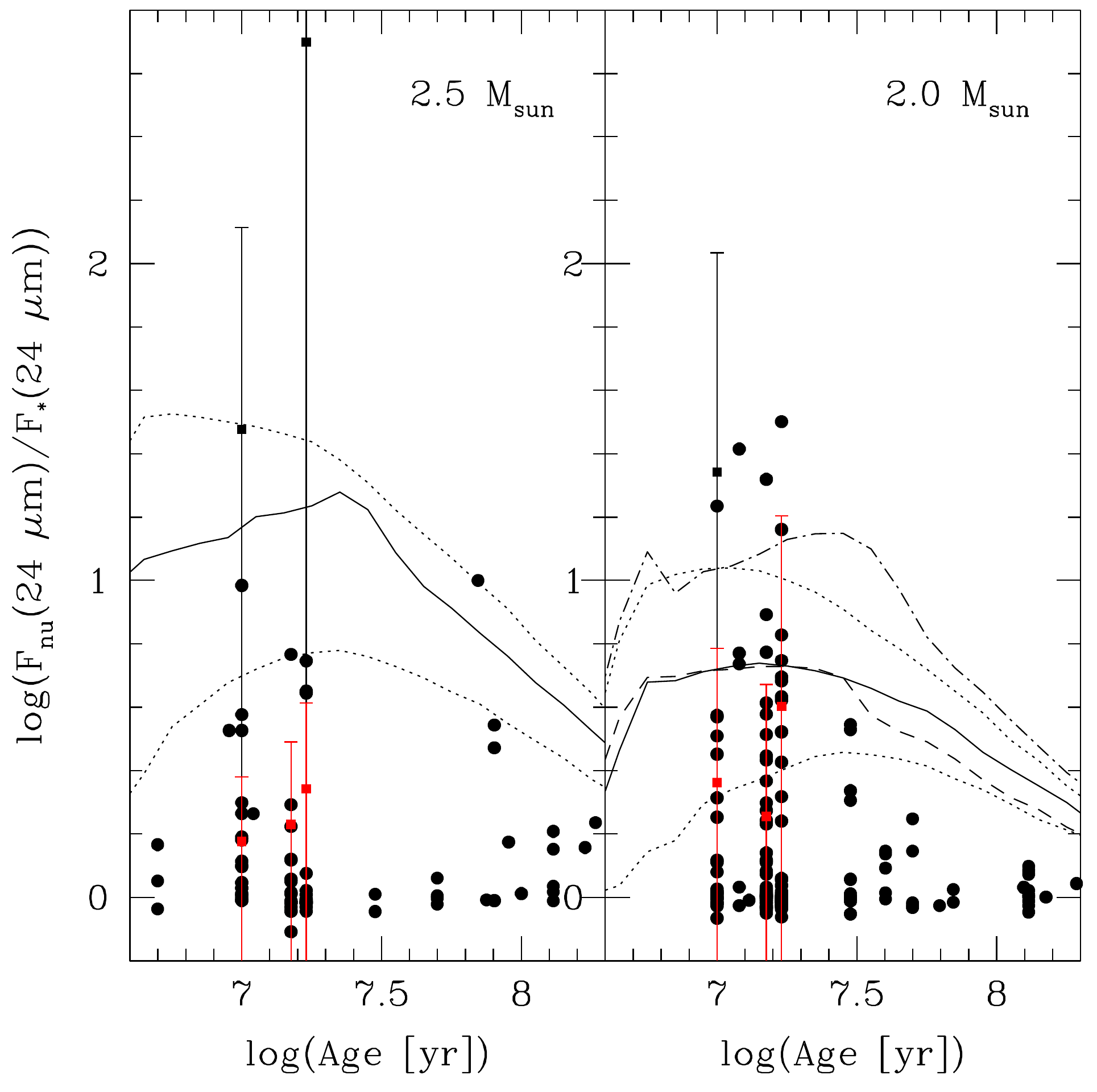}
\figcaption{The 24 $\mu$m flux ratio ($F_{\nu}$(24 $\mu$m)/$F_{*}$(24 $\mu$m)) as a function of stellar age. Our observations of ScoCen are plotted along with data from several young clusters and moving groups in the published literature (See Table 4). Objects with near-infrared (e.g., IRAC) and MIPS 24 $\mu$m excesses are plotted using asterisks; objects with only MIPS 24 $\mu$m excesses are plotted using solid circles. The solid line shows the evolution of thermal emission from dust generated via collisional cascade at distances 30 - 150 AU from the central star \citet(kenyon08) from a typical disk; for 1 $M_{\sun}$ stars, the model corresponds to a Minimum Mass Solar Nebula. The dotted lines show models for disks with masses that are a factor of three higher and lower. The dashed line and the dashed-dotted line show \cite{kenyon10} models for disks with weak planetesimals and an initial planetesimal surface density, $\Sigma$ $\propto$ $a^{-1}$ rather than $a^{-3/2}$. Objects with near- and mid-infrared excesses may be accreting, primordial disks (e.g., HD 100546). Black squares and error bars show mean and standard deviation of ScoCen sample while black red squares and error bars show mean and standard deviation if primordial disks are removed. 
\label{fig:fr24vage}}
\end{figure}

\begin{figure}
\figurenum{7}
\plotone{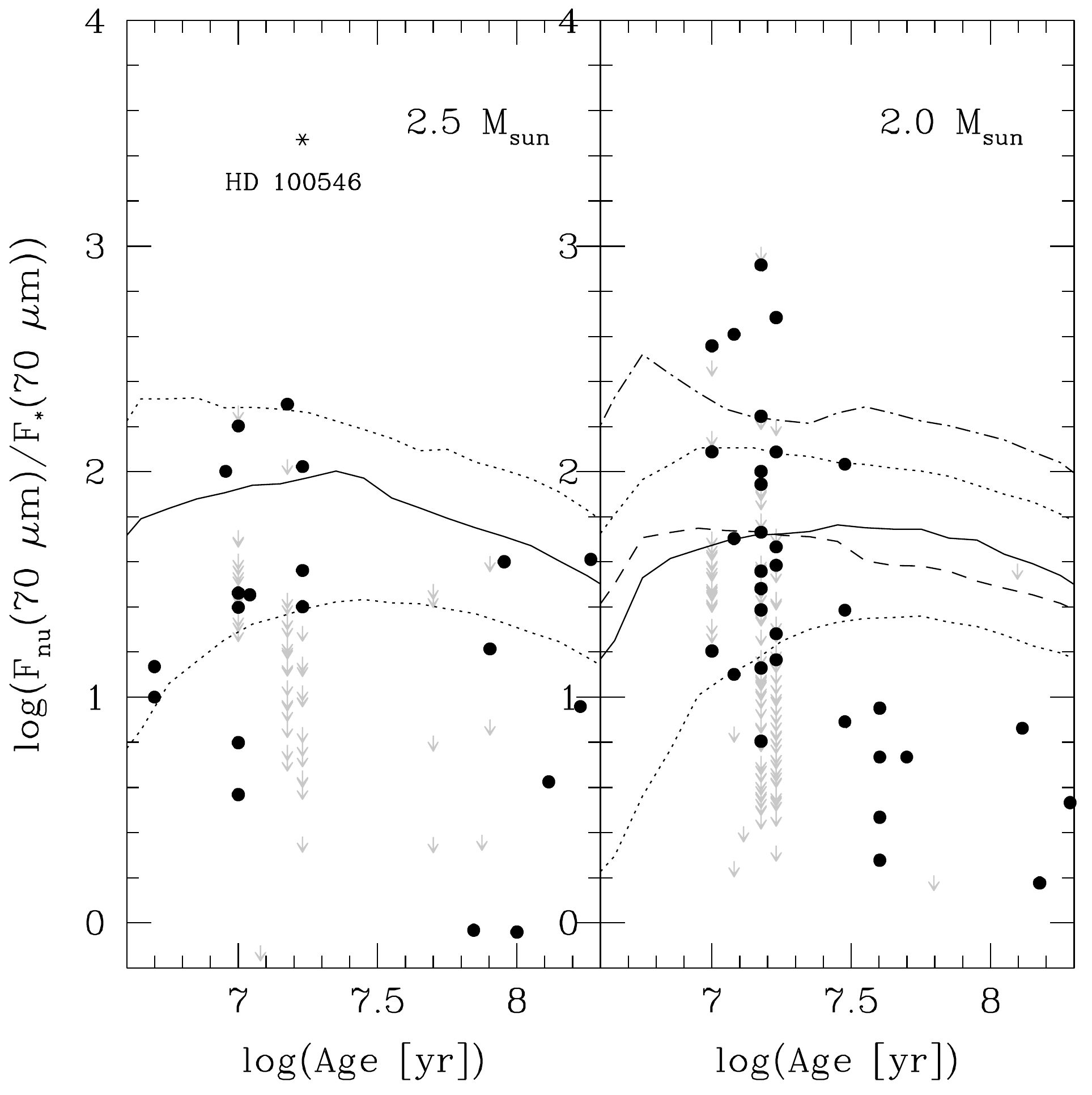}
\figcaption{Same as Figure~\ref{fig:fr24vage} for the 70 $\mu$m flux ratio, ($F_{\nu}$(70 $\mu$m)/$F_{*}$(70 $\mu$m)). Objects that are not detected with MIPS at 70 $\mu$m are plotted as gray upper limit symbols. The solid line shows the evolution of 70 $\mu$m emission from material generated via collisional cascade at distance 30 - 150 AU from the central star \citep{kenyon08} from a typical disk; for 1 $M_{\sun}$ stars, the model corresponds to a Minimum Mass Solar Nebula. The dotted lines show models for disks with masses that are a factor of three higher and lower. The dashed line and the dashed-dotted line show \cite{kenyon10} models for disks with weak planetesimals and an initial planetesimal surface density, $\Sigma$ $\propto$ $a^{-1}$ rather than $a^{-3/2}$.
\label{fig:fr70vage}}
\end{figure}

We plot the MIPS 24 $\mu$m and 70 $\mu$m flux ratios, $F_{\nu}$(observed)/$F_{\nu}$(predicted), for nearby, intermediate-mass stars (2.0$\pm$0.2 M$_{\odot}$ and 2.5$\pm$0.2 M$_{\odot}$ stars within 200 pc of the Sun), including our Sco Cen data and other data from the literature (Table \ref{tab:clusters}) in Figures \ref{fig:fr24vage} and \ref{fig:fr70vage}, consistent with our analysis of solar-like systems \citep{chen11}. Stars with accreting, primordial disks are shown with asterisks while those with debris disks are shown with solid circles. We list the disk type and accretion properties of 2.5 and 2.0 $M_{\sun}$ ScoCen stars in Tables \ref{tab:us} and \ref{tab:lcc}. Our current sample of intermediate-mass ScoCen stars only includes a handful of primordial disks, unlike our previously-studied sample of ScoCen solar-like stars; therefore, inclusions or exclusion of primordial systems has a smaller effect on our current analysis. In general, we find that the 24 $\mu$m excess ratio associated with 2.0-2.5 $M_{\sun}$ stars is consistent with \cite{kenyon04,kenyon08,kenyon10} expectations; however, the observed 24 $\mu$m excess around 2.5 $M_{\sun}$ stars is somewhat lower than predicted. In \cite{chen11}, we hypothesized that 1.0 $M_{\sun}$ stars possessed larger than expected 24 $\mu$m excesses because the inner radii of the disks may have been smaller than 30 AU. We quantify trends in disk properties as a function of stellar mass and describe possible scenarios explaining the observations in the Section 6.

\begin{figure}
\figurenum{8}
\plottwo{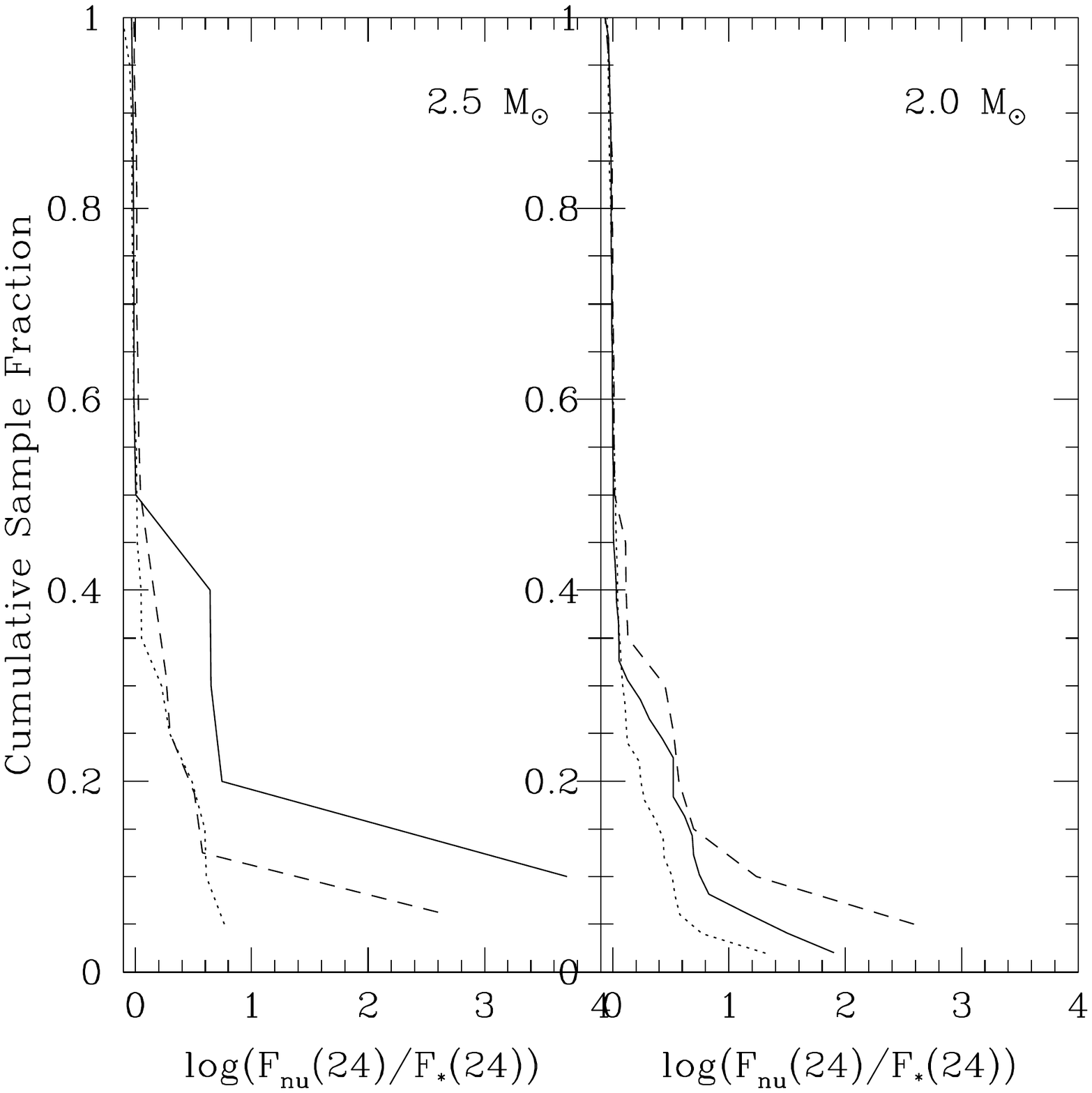}{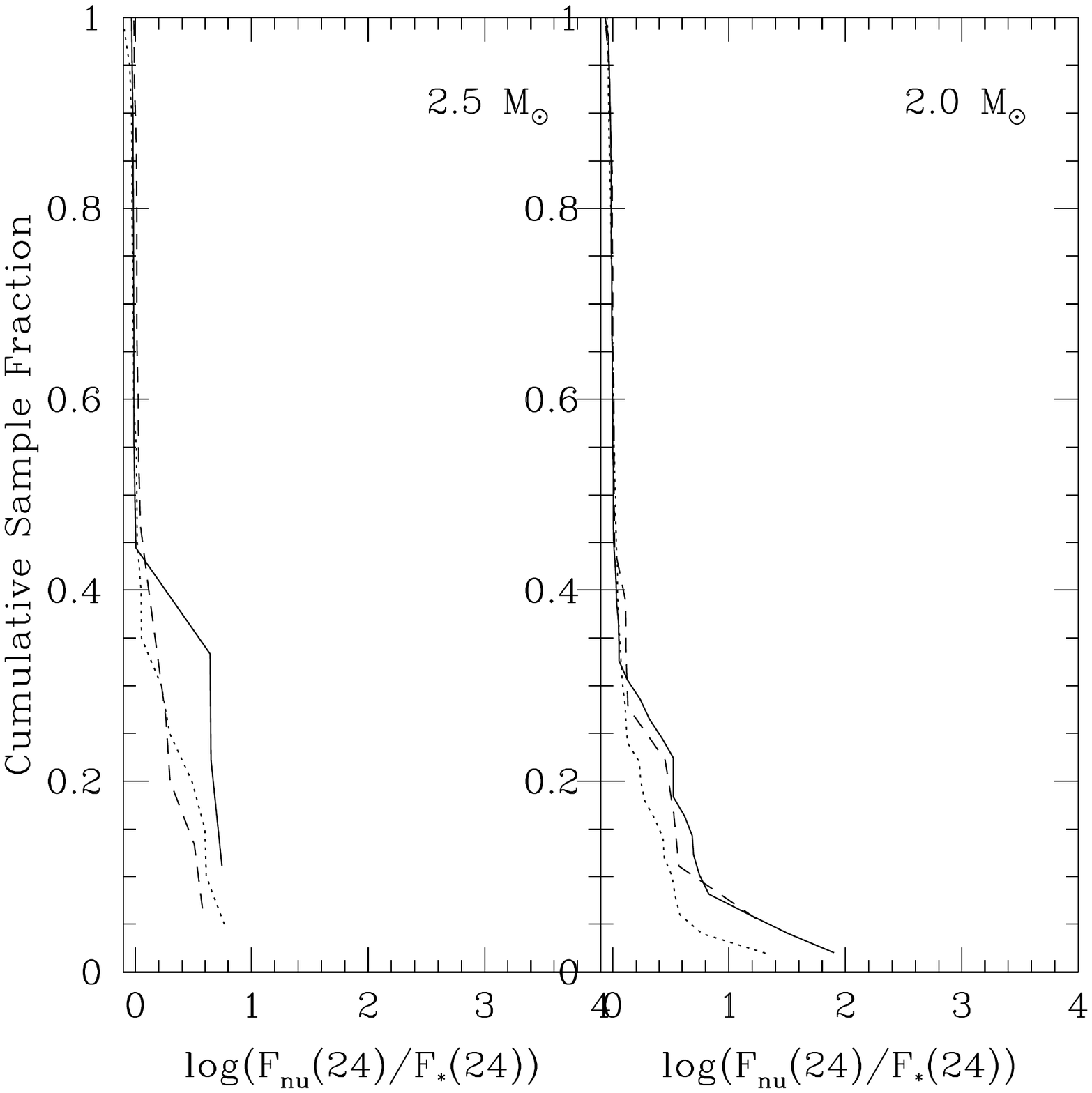}
\figcaption{(a) The cumulative distribution functions of $F_{\nu}$(24 $\mu$m)/$F_{*}$(24 $\mu$m) for 2.5 and 1.0 $M_{\sun}$ stars in US (dashed line), UCL (dotted line), and LCC (solid line) including both "primordial" and "debris" disks. (b) Same as (a) with "primordial" disks removed.
\label{fig:subgrpcd}}
\end{figure}

Since the subgroups of ScoCen possess slightly different ages, we searched for differences in the disk demographics between subgroups. We plot the cumulative disk fractions for 2.0 and 2.5 $M_{\sun}$ stars in US, UCL, and LCC as a function of $F_{\nu}$(24 $\mu$m)/$F_{*}$(24 $\mu$m) (Figure \ref{fig:subgrpcd}). \cite{currie08} searched for statistical trends in similar measurements of B-, A-, and F-type stars in Orion OB1b ($\sim$5 Myr), Orion OB1a ($\sim$10 Myr), and ScoCen ($\sim$10-20 Myr). They performed a Wilcoxon rank sum analysis on measurements of [24] - [24]$_{*}$ in Orion OB1a and ScoCen compared to Orion OB1b. For two samples, the Wilcoxon rank sum test determines the equality or inequality of two distributions by computing the mean rank of one distribution in a combined sample of both distributions. \cite{currie08} compute negative Z parameters and small probabilities, indicating that the mean 24 $\mu$m excess for stars measured in Orion OB1a and ScoCen were statistically smaller than that of Orion OB1b. We performed the Wilcoxon rank sum test on our sample of ScoCen intermediate mass star $F_{\nu}$(24 $\mu$m)/$F_{*}$(24 $\mu$m) measurements, comparing each subgroup with US and UCL for 1.0 and 1.5 $M_{\sun}$ stars both including and excluding primordial disks (see Table \ref{tab:statistics}). For both 2.0 and 2.5 $M_{\sun}$ stars, we find that the Wilcoxon rank sum test yields large probabilities ($>$4\%) when comparing US to UCL and LCC, indicating that the excesses around UCL and LCC are not statistically indistinguishable from those around US.

To better understand the $F_{\nu}$(24 $\mu$m)/$F_{*}$(24 $\mu$m) distributions, we calculated the first quartile, median, and third quartile $F_{\nu}$(24 $\mu$m)/$F_{*}$(24 $\mu$m) for 2.5 and 2.0 M$_{\sun}$ stars in US, UCL, and LCC (see Table \ref{tab:statistics}). Typical measurement uncertainties for F$_{\nu}$(24 $\mu$m)/$F_{*}$(24 $\mu$m) in	US, UCL,	and LCC are 0.07, 0.05, and 0.05, respectively, suggesting that the median $F_{\nu}$(24 $\mu$m)/$F_{*}$(24 $\mu$m)	for each subgroup is consistent with	a bare photosphere. We calculate the mean and standard deviation of F$_{\nu}$(24 $\mu$m)/$F_{*}$(24 $\mu$m) (see Table \ref{tab:statistics}) and overlay these values in our diagram showing excess trends as a function of age (Figure \ref{fig:fr24vage}). For	 both	2.5 and 2.0 M$_{\sun}$ stars, the $F_{\nu}$(24 $\mu$m)/$F_{*}$(24 $\mu$m) standard deviation is	very	large, making it difficult to determine whether any robust evolutionary trend exists from the mean or the median alone. However, the first quartile values probe the strength of the 24 $\mu$m excess systems and indicate increased 24 $\mu$m excess for 2.0 and 2.5 M$_{\sun}$ stars, regardless of whether all disks are included or primordial disks are excluded from the analysis. Unfortunately, we detect only $\sim$10\% of our objects at 70 $\mu$m. The detected objects possess excesses that are consistent with the expectations from KB08; however, the majority of the systems possess 3$\sigma$ flux upper limits that are consistent with the models.

\section{Dust Grain Properties}

We compare the grain properties for detected ScoCen debris disks with those measured for debris disks in mature planetary systems. We estimate typical measured fractional infrared luminosities (2$\times$10$^{-6}$ $<$ $L_{IR}/L_{*}$ $<$ 4$\times$10$^{-2}$ and MIPS 24 and 70 $\mu$m blackbody color temperatures (40 K $<$ $T_{gr}$ $<$ 300 K) around B- and A-type stars at $\sim$10-20 Myr. \cite{su06} estimate typical fractional infrared luminosities (3$\times$10$^{-6}$ $<$ $L_{IR}/L_{*}$ $<$ 1$\times$10$^{-3}$) and color temperatures (60 K $<$ $T_{gr}$ $< $ 250 K) for debris disks around 1 Myr to 1 Gyr B6-A7-type stars. Our sample contains nine disks with somewhat higher fractional infrared luminosities ($L_{IR}/L_{*}$ $>$ 1$\times$10$^{-3}$) and warm terrestrial temperature debris ($T_{gr}$ $>$ 200 K), consistent with the younger ages of these systems. 

We estimate the minimum dust grain size, $a_{min}$, assuming that radiation pressure removes the smallest grains if $\beta$ (=F$_{rad}$/F$_{grav}$) $>$ 0.5.
\begin{equation}
  {a_{min} > \frac{6L_{*}<Q_{pr}(a)>}{16 \pi GM_{*}c \rho_{s}}}
  \label{eqn-amin}
\end{equation}
\citep{artymowicz88} where $L_{*}$ and $M_{*}$ are the stellar luminosity and mass, $<Q_{pr}(a)>$ (=($\int F_{\lambda}d\lambda$)$^{-1}$ $\int Q_{pr}(a,\lambda)F_{\lambda}d \lambda$) is the radiation pressure coupling coefficient, and $\rho_{s}$ is the density of an individual dust grain. We estimate stellar masses for UCL/LCC members using our estimated stellar effective temperatures and a 15 Myr isochrone \citep{bertelli94,dotter08}. Infrared spectroscopy of T Tauri disks suggest that the bulk of the dust in these systems is probably amorphous olivine \citep{watson09}; therefore, we use optical constants measured for amorphous olivine ($\rho_{s}$ = 3.71 g cm$^{-3}$; \cite{dorschner95}) to estimate the radiation pressure coupling coefficient; for simplicity, we assume that the grains are spherical. In general, the estimated minimum grain sizes are too small for the grains to behave as simple blackbodies ($Q_{abs}$ $\propto$ constant) and too large for the grains to behave in the small grain approximation (2$\pi a$ $<<$ $\lambda$).

We estimate the grain distance of the dust from the grain temperature, $T_{gr}$, assuming that the dust particles are in radiative equilibrium, and possess an average grain size, $<a> = 5/3$ a$_{min}$, expected if the grains are in collisional equilibrium. Dust grains, $D$, in radiative equilibrium with a star are located at a distance, $D$  
\begin{equation}
\left(\frac{R_{*}}{D}\right)^{2}\int_0^\infty Q_{abs}(\nu)B_{\nu}(T_{*})d\nu = 4\int_0^\infty Q_{abs}B_{\nu}(T_{gr})d\nu
\end{equation} 
where $Q_{abs}$ is the absorption coefficient for the dust grains. We estimate dust distances using optical constants measured for amorphous olivine and calculate absorption coefficients assuming that the grains are spherical, with radius, $<a>$. Our data are consistent with the presence of dust located in rings at Kuiper-Belt-like distances, suggesting that the average ScoCen debris disk is a massive analog to our Kuiper Belt at an age $<$20 Myr.

We estimate the minimum mass of infrared-emitting dust grains, assuming that the grains have a radius, $<a>$; if the grains are larger, then our estimate is a lower bound. If we assume a thin shell of dust at distance, $D$, from the central star, and if the grains are spheres of radius, $<a>$, and if the cross section of the grains is equal to their geometric cross section, then the mass of dust is
\begin{equation}
M_d \ge \frac{16}{3}\pi \frac{L_{IR}}{L_{*}}\rho_{s}D^{2}<a>
\end{equation}
where $L_{IR}$ is the luminosity of the dust. The bulk of the dust mass is expected to be located in larger grains.

We estimate the minimum mass in parent bodies assuming that the disk is Poynting--Robertson (PR) drag dominated and in steady state. We hypothesize that each system possesses at least as much mass in parent bodies today as that which would have been destroyed if the system were in steady state during the lifetime of the star. If MPB denotes the mass in parent bodies, then we may write
\begin{equation}
M_{PB} \geq  \frac{4 L_{IR} t_{age}}{c^2}
\end{equation}
\citep{chen01}. If the disk is dominated by collisions, as is probably the case, then this estimate will be a lower bound.

\section{Stellar Mass Dependence}
\begin{figure}
\figurenum{4}
\plotone{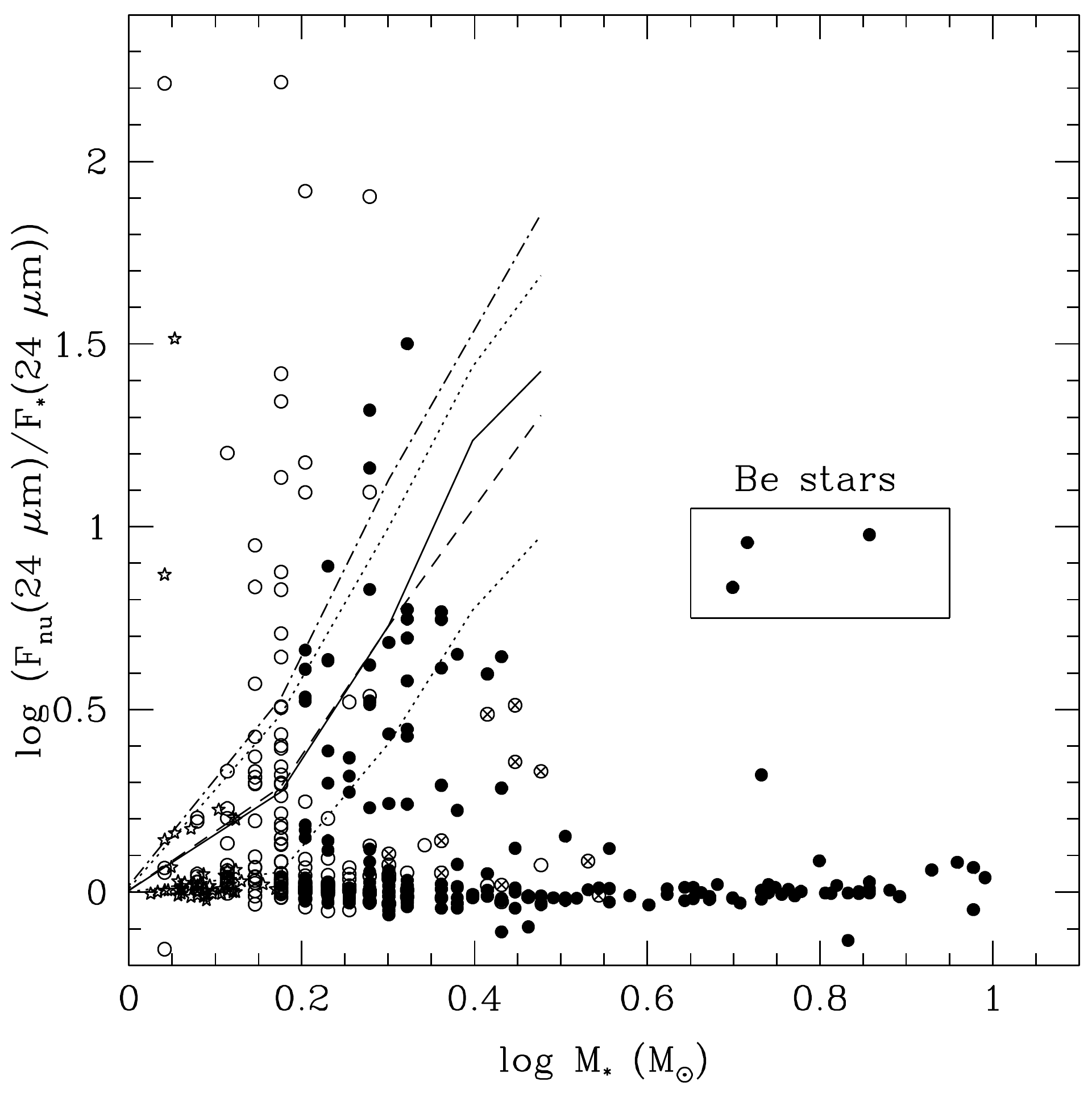}
\figcaption{The 24 $\mu$m flux ratio ($F_{\nu}$(24 $\mu$m)/$F_{*}$(24 $\mu$m)) plotted as a function of stellar mass. B- and A-type stars in this survey are shown with solid circles while F- and G-type stars from \cite{chen11} are shown with open circles. The solid line shows the 24 $\mu$m flux ratios predicted in the \cite{kenyon08} model for typical disks with ages $\sim$10 - 20 Myr. The dotted lines show models for disks with masses that are a factor of three higher and lower than typical disks. The dashed line and the dashed-dotted line show \cite{kenyon10} models for disks with weak planetesimals and an initial planetesimal surface density, $\Sigma$ $\propto$ $a^{-1}$ rather than $a^{-3/2}$. High mass stars with large infrared excess are Be stars. 
\label{fig:r24}}
\end{figure}

\begin{figure}
\figurenum{5}
\plotone{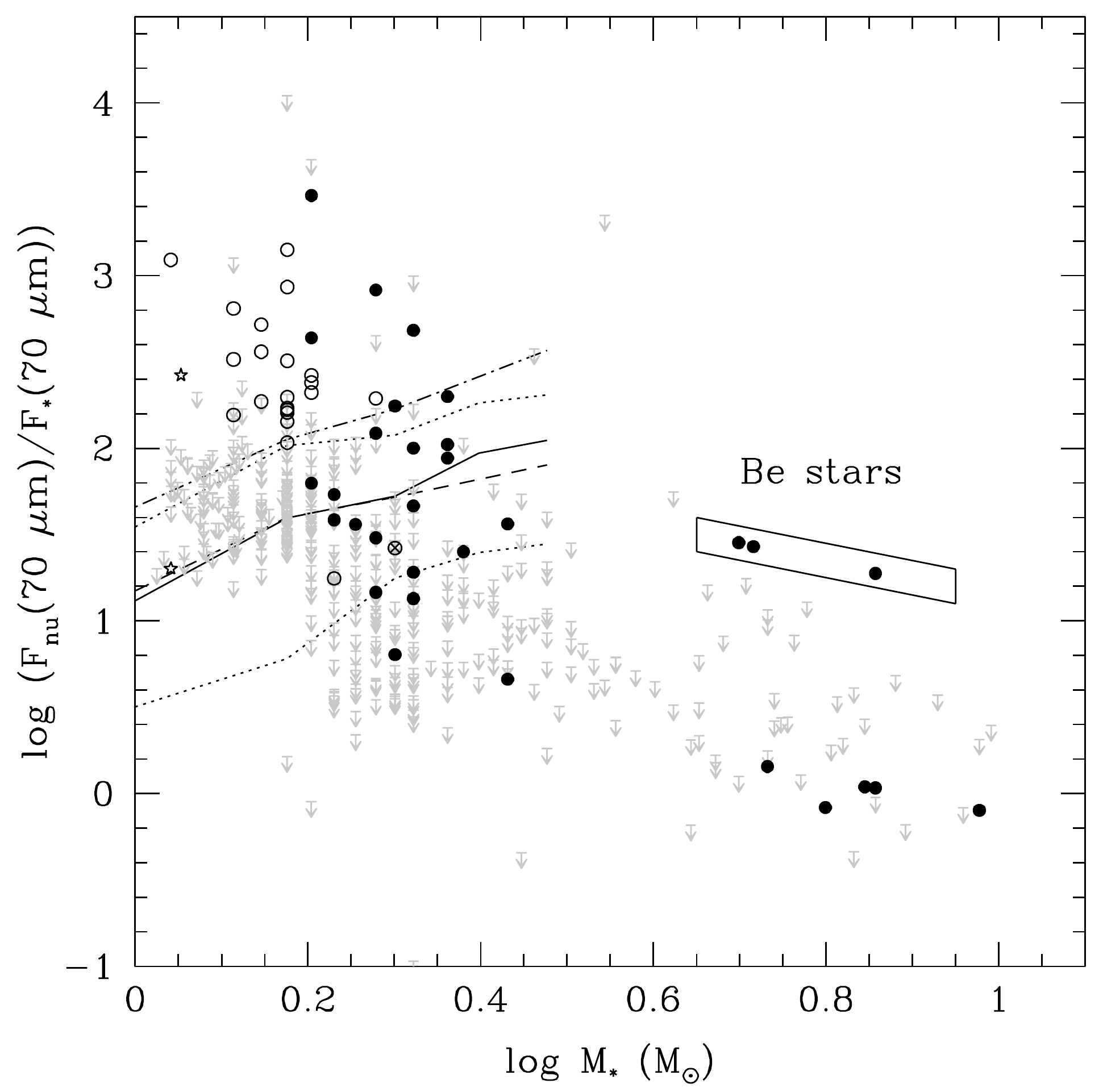}
\figcaption{Same as Figure~\ref{fig:r24} for the 70 $\mu$m flux ratio, ($F_{\nu}$(70 $\mu$m)/$F_{*}$(70 $\mu$m)). 
\label{fig:r70}}
\end{figure}

We combine the data from our survey of 215 ScoCen B- and A-type stars with that from the survey of \cite{chen11} of 181 ScoCen F- and G-type stars to search for trends in disk properties as a function of stellar mass. We find that (1) few stars with masses $>$3 $M_{\sun}$ possess 24 $\mu$m and/or 70 $\mu$m excess and (2) stars with masses 1.0-1.75 $M_{\sun}$ possess larger first quartile and median $F_{\nu}$(24 $\mu$m)/$F_{*}$(24 $\mu$m) and $F_{\nu}$(70 $\mu$m)/$F_{*}$(70 $\mu$m) than stars with masses 1.75-3.0 $M_{\sun}$ (see Figures \ref{fig:r24} and \ref{fig:r70}). The first quartile $F_{\nu}$(24 $\mu$m)/$F_{*}$(24 $\mu$m) values for 1.0-1.75 $M_{\sun}$ and 1.75-3.0 $M_{\sun}$ stars are 1.59 and 1.32, respectively. Comparing the 24 $\mu$m flux ratio measurements of the low mass stars with those of the high mass stars, the Wilcoxon-Rank Sum test yields a Z-parameter of 2.96 and a probability of 0.2\%, indicating that the 1.0-1.75 $M_{\sun}$ stars possess a higher median $F_{\nu}$(24 $\mu$m)/$F_{*}$(24 $\mu$m) value than the 1.75-3.0 $M_{\sun}$ stars. The first quartile $F_{\nu}$(70 $\mu$m)/$F_{*}$(70 $\mu$m) values for 1.0-1.75 $M_{\sun}$ and 1.75-3.0 $M_{\sun}$ stars are 520 and 175, respectively. Comparing the 70 $\mu$m flux ratio measurements of the low mass stars with those of the high mass stars, the Wilcoxon-Rank Sum test yields a Z-parameter of 3.62 and a probability of 0.01\%, indicating that the 1.0-1.75 $M_{\sun}$ stars possess a higher median $F_{\nu}$(70 $\mu$m)/$F_{*}$(70 $\mu$m) value than the 1.75-3.0 $M_{\sun}$ stars.  While our shallow MIPS observations are sensitive to stellar photospheres at 24 $\mu$m, they are not sensitive to stellar photospheres at 70 $\mu$m. Since our $F_{\nu}$(70 $\mu$m)/$F_{*}$(70 $\mu$m) measurements are more sensitive for higher mass stars, our $F_{\nu}$(70 $\mu$m)/$F_{*}$(70 $\mu$m) statistics may be biased.

\begin{figure}
\figurenum{9}
\plotone{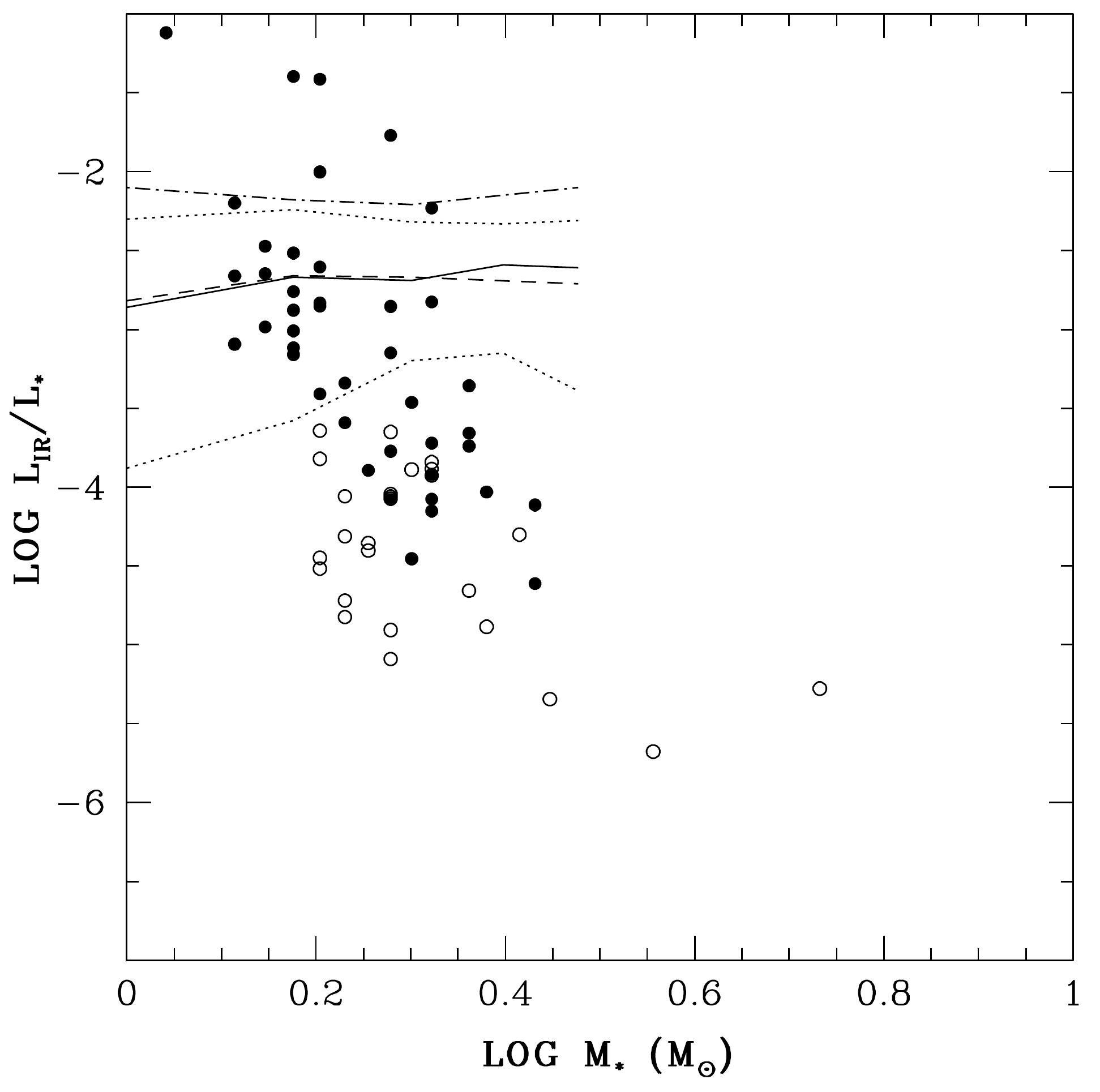}
\figcaption{Fractional infrared luminosity ($L_{IR}/L_{*}$) of excess stars plotted as a function of stellar mass. Stars with infrared excess detected at both 24 and 70 $\mu$m are shown with solid circles while those that are only detected at 24 $\mu$m are shown with open circles. The solid line shows the fractional infrared luminosities predicted in the \cite{kenyon08} model for typical disks in which disk mass scales with stellar mass. The dotted lines show models for disks with masses that are a factor of three higher and lower. The dashed line and the dashed-dotted line show \cite{kenyon10} models for disks with weak planetesimals and an initial planetesimal surface density, $\Sigma$ $\propto$ $a^{-1}$ rather than $a^{-3/2}$. Be stars with infrared excesses have been removed from the sample. 
\label{fig:fd}}
\end{figure}

\begin{figure}
\figurenum{10}
\plotone{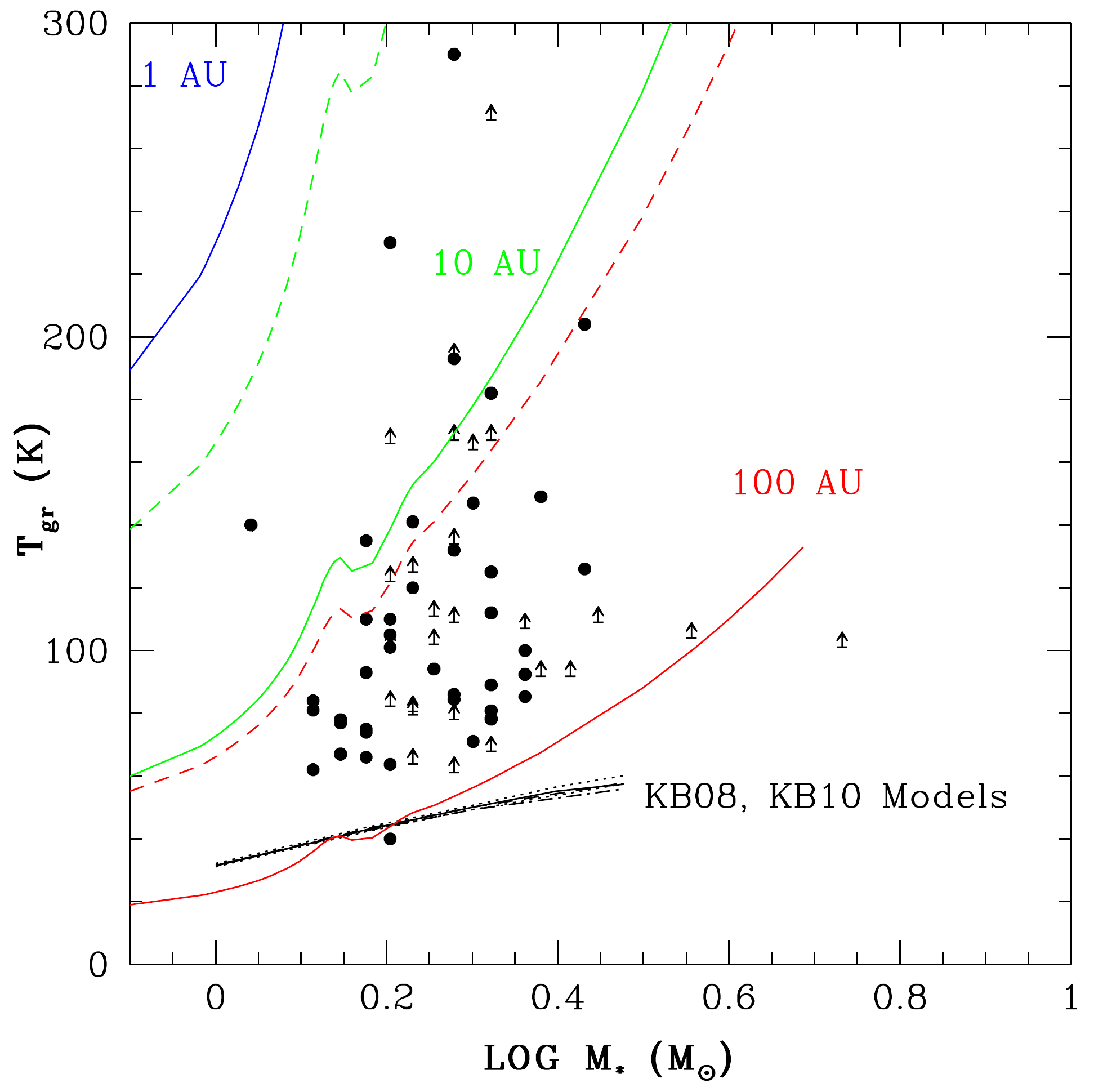}
\figcaption{Grain temperatures ($T_{gr}$) of excess stars plotted as a function of stellar mass. The blue, green, and red solid lines show grain temperatures expected for black body grains at 1 AU, 10 AU, and 100 AU from the central star while the green and red dashed lines show grain temperatures expected for small grains (2$\pi a$ $<<$ $\lambda$) at the same distances. The solid black line shows the grain temperatures predicted in the \cite{kenyon08} model for typical disks in which disk mass scales with stellar mass. The dashed lines show models for disks with masses that are a factor of three higher and lower. The dashed line and the dashed-dotted line show \cite{kenyon10} models for disks with weak planetesimals and an initial planetesimal surface density, $\Sigma$ $\propto$ $a^{-1}$ rather than $a^{-3/2}$. Be stars with infrared excesses have been removed from the sample. 
\label{fig:tgr}}
\end{figure}

\begin{figure}
\figurenum{11}
\plotone{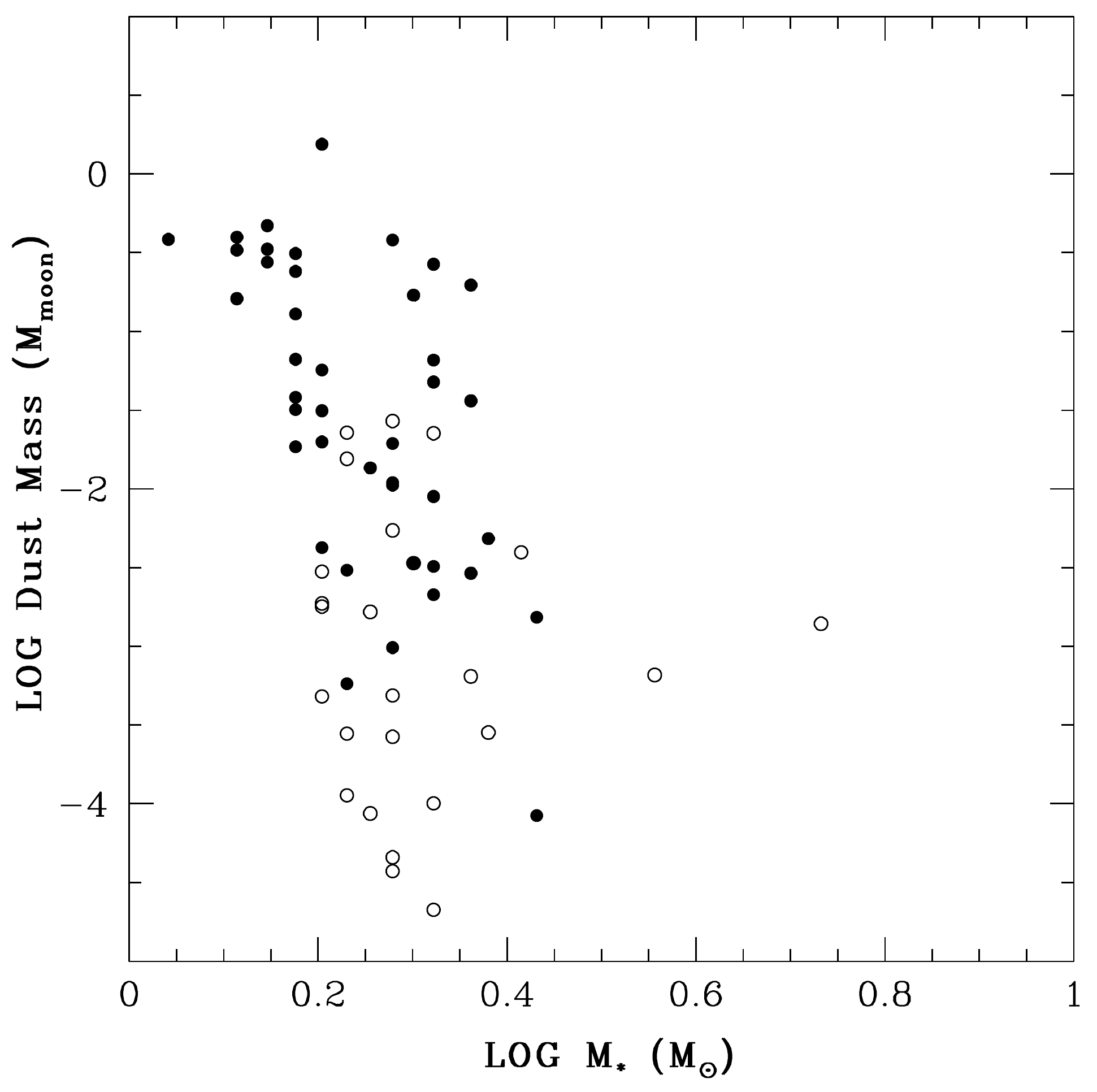}
\figcaption{Estimated dust mass ($M_{d}$) of excess stars plotted as a function of stellar mass.  
\label{fig:mdust}}
\end{figure}

\begin{figure}
\figurenum{12}
\plotone{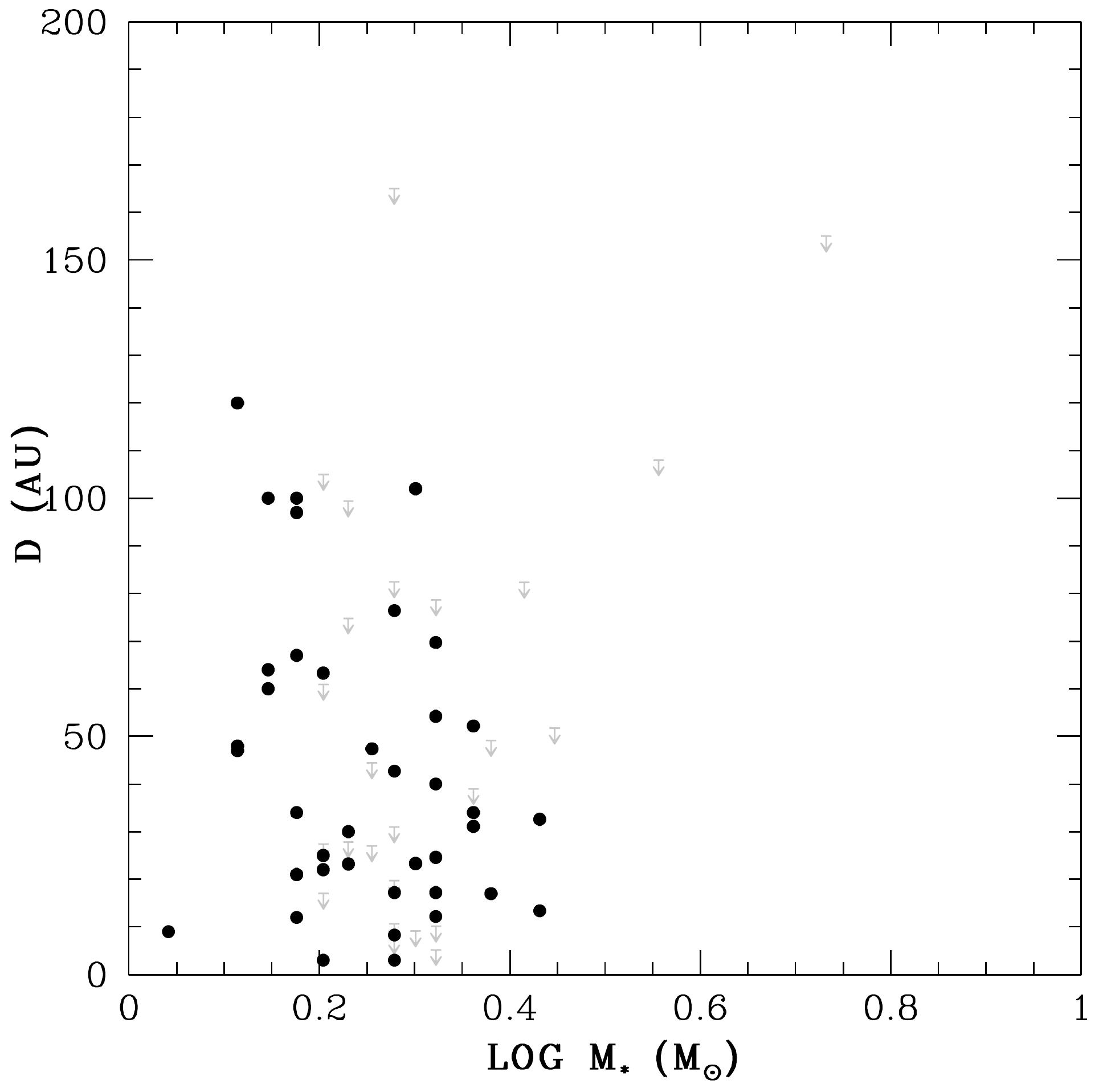}
\figcaption{Estimated dust distance ($D$) of excess stars plotted as a function of stellar mass. Stars with infrared excess detected at both 24 and 70 $\mu$m are shown with solid circles while those that are only detected at 24 $\mu$m are shown with open circles. 
\label{fig:ddust}}
\end{figure}

Since debris disks are typically characterized by their the fractional infrared luminosities and dust grain temperatures, we also searched for trends in these quantities as a function of stellar mass. We find that $L_{IR}/L_{*}$ is mass dependent with 1.0-1.75 $M_{\sun}$ stars possessing higher $L_{IR}/L_{*}$ than 1.75-3.0 $M_{\sun}$ stars (see Figure \ref{fig:fd}). The first quartile $L_{IR}$/$L_{*}$ values for 1.0-1.75 $M_{\sun}$ and 1.75-3.0 $M_{\sun}$ stars are 0.003 and 0.0002, respectively. Comparing the fractional infrared luminosity measurements of the low mass stars with those of the high mass stars, the Wilcoxon-Rank Sum test yields a Z-parameter of 3.71 and a probability of 0.01\%, indicating that the 1.0-1.75 $M_{\sun}$ stars possess a higher median $F_{\nu}$(24 $\mu$m)/$F_{*}$(24 $\mu$m) value than the 1.75-3.0 $M_{\sun}$ stars. If the dust grains are optically thin, then our results may indicate that the dust mass in micron-sized grains decreases as a function of stellar mass. Measurements of the thermal continuum at submillimeter and/or millmeter wavelengths, where the dust is more likely to be optically thin, are needed to determine definitively whether dust mass declines with stellar mass. We do not find any dependence of $T_{gr}$ on stellar mass (see Figure \ref{fig:tgr}); however, we note that our temperatures are only estimated from our MIPS 24 $\mu$m and 70 $\mu$m photometry. Measurements of the infrared excess at other wavelengths are needed to better characterize the dust temperature in detail. For a more direct comparison with planet formation models, we also plot our estimated dust masses, $M_{d}$, and estimated dust distances, $D$, as a function of stellar mass (see Figure \ref{fig:mdust} and Figure \ref{fig:ddust}). We note that (1) the estimated dust distance depends sensitively on assumed grain properties and that observations of resolved disks indicate that similarly estimated distances may be inconsistent with observations by as much as a factor of two and (2) the estimated dust mass only represents the estimated mass in micron-sized grains observed at mid- to far-infrared wavelengths. Our estimated dust distances indicate that the cold planetesimal belts in ScoCen are too distant to be influenced by giant planets on circular orbits that are predicted to have formed via traditional core-accretion models. Studies of the location of the snow-line in giant planet forming disks indicate that giant planets form at $\sim$6-10 AU and $\sim$8-20 AU around 1 and 3 $M_{\sun}$ stars, respectively \citep{kennedy08}.

In the \cite{kenyon04,kenyon08,kenyon10} models, the mass of the central star plays a critical role in determining the evolution of the disk. Since the models assume that the mass in initial parent bodies is proportional to stellar mass, the number of small grains generated via collisional cascade is expected to be larger around higher mass stars. The dust in debris disks around higher mass stars is also expected to be warmer than that around lower mass stars because all of the disks are assumed to possess the same inner and outer radii and the higher mass stars are more luminous and therefore more effectively heat circumstellar dust. In Figures \ref{fig:r24} and \ref{fig:r70}, we overplot the \cite{kenyon08} $F_{\nu}$(24 $\mu$m)/$F_{*}$(24 $\mu$m) and $F_{\nu}$(70 $\mu$m)/$F_{*}$(70 $\mu$m) predictions as a function of stellar mass for 11, 15, and 17 Myr old stars with initial disk masses similar to the MMSN and disks with masses 1/3 lower and 3 times higher and \cite{kenyon10} predictions for disks with weak planetesimals or a surface density profile, $\Sigma$ $\propto$ $r^{-1}$ rather than $r^{-3/2}$. We estimate the fraction of stars with excesses above the \cite{kenyon08} model predictions, consistent with the \cite{kenyon08} model predictions, and below the \cite{kenyon08} model predictions for 1.0-1.75 and 1.75-3.0 $M_{\sun}$ stars. For 1.00-1.75 $M_{\sun}$ stars, we estimate that 23/190 (12$^{+3}_{-2}$\%), 82/190 (43$^{+4}_{-3}$\%), and 85/190 (45$^{+4}_{-4}$\%) of systems possess 24 $\mu$m excesses above, consistent with, and below the models. For 1.75-3.00 $M_{\sun}$ stars, we estimate that 5/171 (3$^{+2}_{-1}$\%), 22/171 (13$^{+3}_{-2}$\%), and 144/171 (84$^{+2}_{-3}$\%) of systems possess 24 $\mu$m excesses above, consistent with, and below the models, suggesting that the infrared excesses associated with 1.75-3.0 $M_{\sun}$ is disproportionately low compared with the models. Our estimates for the dust grain temperature around ScoCen stars is significantly higher than that predicted for 10-20 Myr old stars by \citep{kenyon08}. The discrepancy between the observed and \cite{kenyon08} predicted dust grain temperature is probably the result of the artificial 30-150 AU boundary conditions assumed by the models. A similar discrepancy has been noted for debris disks around older, A-type stars and has been used to argue for the presence of narrow debris belts \citep{kennedy10}.

\subsection{Parent Body Mass}
Current models for debris disks assume that the radiation pressure blow-out limit sets the minimum grain size, suggesting that the minimum grain size around higher mass stars is larger than that around lower mass stars. One possible explanation for the smaller infrared excesses around B- and A-type stars is that all systems possess the same parent body mass; however, the higher radiation pressure of higher mass stars disproportionately removes the smallest grains which radiate the bulk of the infrared excess. To determine whether blow-out size effects can account for the discrepancies in our observations, we estimate the increased surface area in small grains generated by lowering the minimum grain size. For dust particles in collisional equilibrium, the surface area in particles, $\sigma_{tot}$ $\propto$ $a_{min}^{-0.5}$ \citep{kennedy10}. Therefore lowering the blowout size from $\sim$10 $\mu$m around a 25 $L_{\sun}$, 2.1 $M_{\sun}$ to $\sim$1 $\mu$m around a 1 $L_{\sun}$, 1 $M_{\sun}$ star is expected to decrease the dust surface area and therefore the fractional infrared luminosity by a factor $\sim$3. Since the ratio of the 1st quartile fractional infrared excesses for 1.0-1.75 $M_{\sun}$ and 1.75-3 $M_{\sun}$ stars is 0.003/0.0002 ($\sim$15), significantly larger than based on minimum grain size effects alone, we conclude that radiation pressure blowout is probably not responsible for the observed fractional infrared luminosity dependence on stellar mass. 

\subsection{Stellar Companions}
A large fraction of ScoCen members may be members of binary or multiple systems. Forty B- and A-type and thirty-three F- and G-type ScoCen stars are listed as binary or multiple systems in the Catalog of the Components of Double and Multiple stars (CCDM, \cite{dommanget02}. More recently, \cite{kouwenhoven05} have carried out a near-infrared adaptive optics search (using VLT ADONIS) for companions around 199 B- and A-type members of ScoCen; they discovered 41 new candidate companions with angular separations 0.22$\arcsec$ - 12.4$\arcsec$, corresponding to projected separations of 28.6 AU - 1600 AU. A follow-up demographic study of ScoCen multiplicity, taking into account all available observations of ScoCen intermediate-mass, visual, spectroscopic, and astrometric binaries suggests that the ScoCen intermediate-mass binary fraction may be as large as $>$70\% \citep{kouwenhoven07}. For each multiple system in our sample, we list the estimated angular separation(s) between the primary and its known and/or candidate companion(s) in Table~\ref{tab:starprops}.

\begin{figure}
\figurenum{13}
\plotone{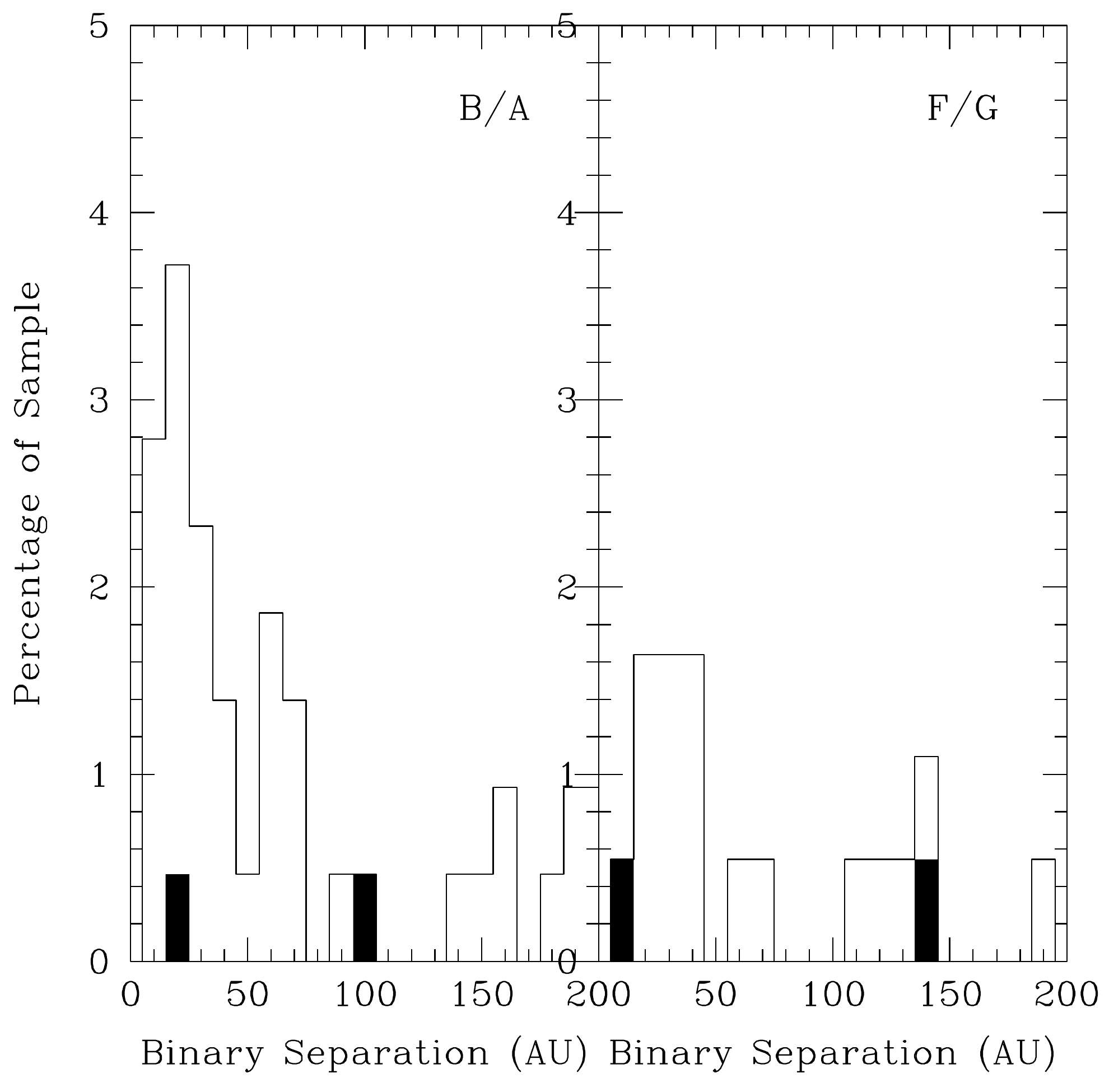}
\figcaption{Histograms show the percentage of stars in (a) our B- and A-type ScoCen sample and (b) the \cite{chen11} F- and G-type ScoCen samples that possess stellar companions at distances 10 - 200 AU (solid line). Overlaid are histograms showing the percentage of binary systems with 24 $\mu$m excesses (filled-in histogram).
\label{fig:binsep}}
\end{figure}

Binary systems have been observed to possess both circumbinary disks that surround both stars and circumstellar disk(s) that surround the primary and/or secondary star(s). Detailed SPH modeling of binary systems suggest that the inner edge of the circumbinary disk is typically located at 1.8$a$ to 3$a$ and that the outer edges of the circumstellar disks are located at $<$0.5$a$, where $a$ is the binary semi-major axis \citep{lubow00}. Therefore, companions may truncate the inner and/or outer edges of planetesimals distributions, modifying the expected properties of collisionally-generated, micron-sized grains (e.g. distance, temperature). To determine whether companions may truncate the outer edges of ScoCen disks, we plotted a histogram showing the percentage of the sample with binary separations 10 - 200 AU for intermediate-mass and solar-mass stars (see Figure \ref{fig:binsep}). We estimate that $\sim$20\% of the B- and A-type stars in our ScoCen sample and $\sim$10\% of the F- and G-type stars in the \cite{chen11} sample possess companions within 200 AU. We hypothesize that the lower excesses associated with intermediate-mass stars may be due to disk truncation; however, we note that the ScoCen F- and G-type members have not been systematically searched for companions. Indeed, we described two new companions to two solar-like ScoCen stars (HIP 72033B and HIP 77520B) for the first time in \cite{chen11}. In Figure \ref{fig:binsep}, we overlaid the distribution of binary systems with 24 $\mu$m excesses; we additionally found that most binary systems do not possess 24 $\mu$m excess.

Two A-type stars (HIP 77150 and HIP 77315) and two F-type stars (HIP 56673/HD 101088, and HIP 63975/HD 113766) apparently possess 24 $\mu$m excesses despite the presence of companions at 10 - 200 AU. For HIP 77315 and HIP 63975, the excesses are detected at 24 and 70 $\mu$m, consistent with the presence of dust at 12.2 AU and 3 AU, respectively, significantly closer to the star than the observed companions at 100 AU and 150 AU, respectively, and consistent with either circumprimary and/or circumsecondary disks. For HIP 77150 and HIP 56673, the excesses are only detected at 24 $\mu$m; therefore, the estimates of grain temperature are lower limits and those of dust distance are upper limits. For HIP 77150, the dust is expected be located at a distance $<$82 AU, consistent with the presence of circumbinary dust around the 25 AU separation binary. For HIP 56673 with a binary separation of 19 AU, the data are more challenging to interpret. Follow-up IRS spectra indicate that the shape of the 5-35 $\mu$m spectrum is photospheric, suggesting that this system does not possess thermal emission from circumstellar dust; however, multi-epoch, high-resolution visual spectra indicate the presence of time-variable H$\alpha$ emission consistent with the presence of an accretion disk around one of the components in this binary system \citep{bitner10}. 

\section{Discussion}
The \emph{Wide-field Infrared Survey Explorer} (WISE, \cite{wright10}) recently completed its mission to map the sky at 3.4, 4.6, 12, and 22 $\mu$m with an angular resolution of 6.1$\arcsec$, 6.4$\arcsec$, 6.5$\arcsec$, and 12.0$\arcsec$, and 5$\sigma$ point source sensitivities $\sim$0.08, 0.11, 1, and 6 mJy. In March 2012, the mission released an all-sky photometric catalog. Since ScoCen members are $\sim$10 - 20 Myr old, their SEDS are expected to be photospheric at 3.4 and 4.6 $\mu$m with possible excesses at 12 and 22 $\mu$m. In 2MASS $K_{s}$, the B- through G-type members possess measured $K_{s}$-band magnitudes typically between 4th and 9th magnitude, suggesting that while the WISE 3.4 and 4.6 $\mu$m photometry is typically saturated, the 2MASS J, H, and $K_{s}$-band photometry is not. Therefore, we prefer to estimate photospheric fluxes at mid-infrared wavelengths from 2MASS data rather than WISE data even though the WISE data are observed at longer wavelengths and would provide more accurate photospheric estimates if the data were not saturated. In addition, we prefer to compare the WISE 12 and 22 $\mu$m brightnesses directly with 2MASS J, H, $K_{s}$ brightnesses to determine whether sources possess long wavelength excesses.

\begin{figure}
\figurenum{14}
\plotone{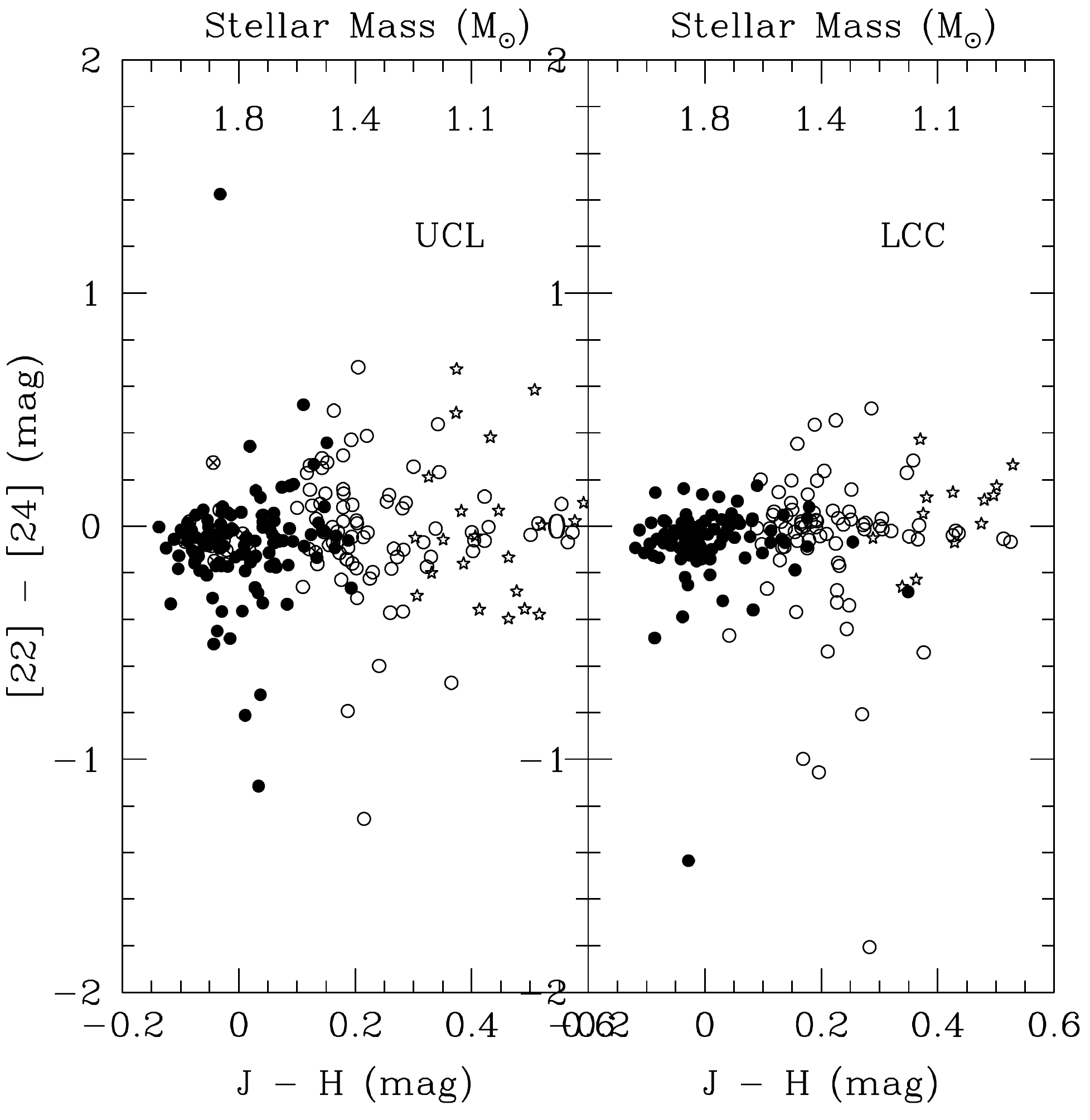}
\figcaption{The WISE [22] - MIPS [24] color plotted as a function of the J - H color for the Upper Centaurus Lupus (UCL) and Lower Centaurus Crux (LCC) subgroups of Sco Cen. Our sample of 209 B- and A-type stars is shown as filled circles; open circles with crosses represent  A-type stars from \citet{su06}; F- and G-type stars from \citet{chen11} are shown as open circles; and the \citet{carpenter08} sample of F- and G-type stars is shown as open stars. 
\label{fig:f2224}}
\end{figure}

Similarly, we expect that our \emph{Spitzer} MIPS 24 $\mu$m photometry observations should supercede the WISE 22 $\mu$m observations because (a) the MIPS 24 $\mu$m beam has an angular size $\sim$6$\arcsec$, half the size of WISE 22 $\mu$m beam, and is therefore less subject to source confusion and (b) our MIPS 24 $\mu$m measurements are more sensitive than the WISE 22 $\mu$m measurements and therefore possess higher signal-to-noise ratios. We compare the WISE 22 $\mu$m and MIPS 24 $\mu$m photometry for all of the ScoCen members that were observed using MIPS (see Figure \ref{fig:f2224}). In general, we find that the 22 and 24 $\mu$m fluxes are well-correlated with a modest dispersion; the UCL and LCC mean [22]-[24] colors are -0.06 mag and -0.04 mag, respectively, with standard deviations of 0.29 mag and 0.30 mag, respectively, suggesting that the WISE 22 $\mu$m data are consistent with the MIPS 24 $\mu$m data to $\sim$30\%. Two stars, HIP 67472 (a Be star in UCL) and PDS 66 (a protoplanetary disk in LCC), possess [22]-[24] $>$1 mag; since these objects possess 24 and 70 $\mu$m excesses, the [22]-[24] colors are probably a measurement of the local SED slope. Three stars in UCL (HIP 80663, HIP 80897, and HIP 81380) and four stars stars in LCC (HIP 53701, HIP 59716, HIP 62677, and HIP 67230) possess [22]-[24] $<$ -1 mag. The majority of the outliers (HIP 53701, HIP 62677, HIP 67230, HIP 80897, HIP 80897, and HIP 81380) are located in regions with bright cirrus; therefore, both MIPS and WISE photometry measurements are probably confused. The MIPS 24 $\mu$m image of HIP 59716 reveals a bright infrared companion at 2.6$\arcsec$ that probably confuses the WISE 22 $\mu$m photometry.

\begin{figure}
\figurenum{15}
\plotone{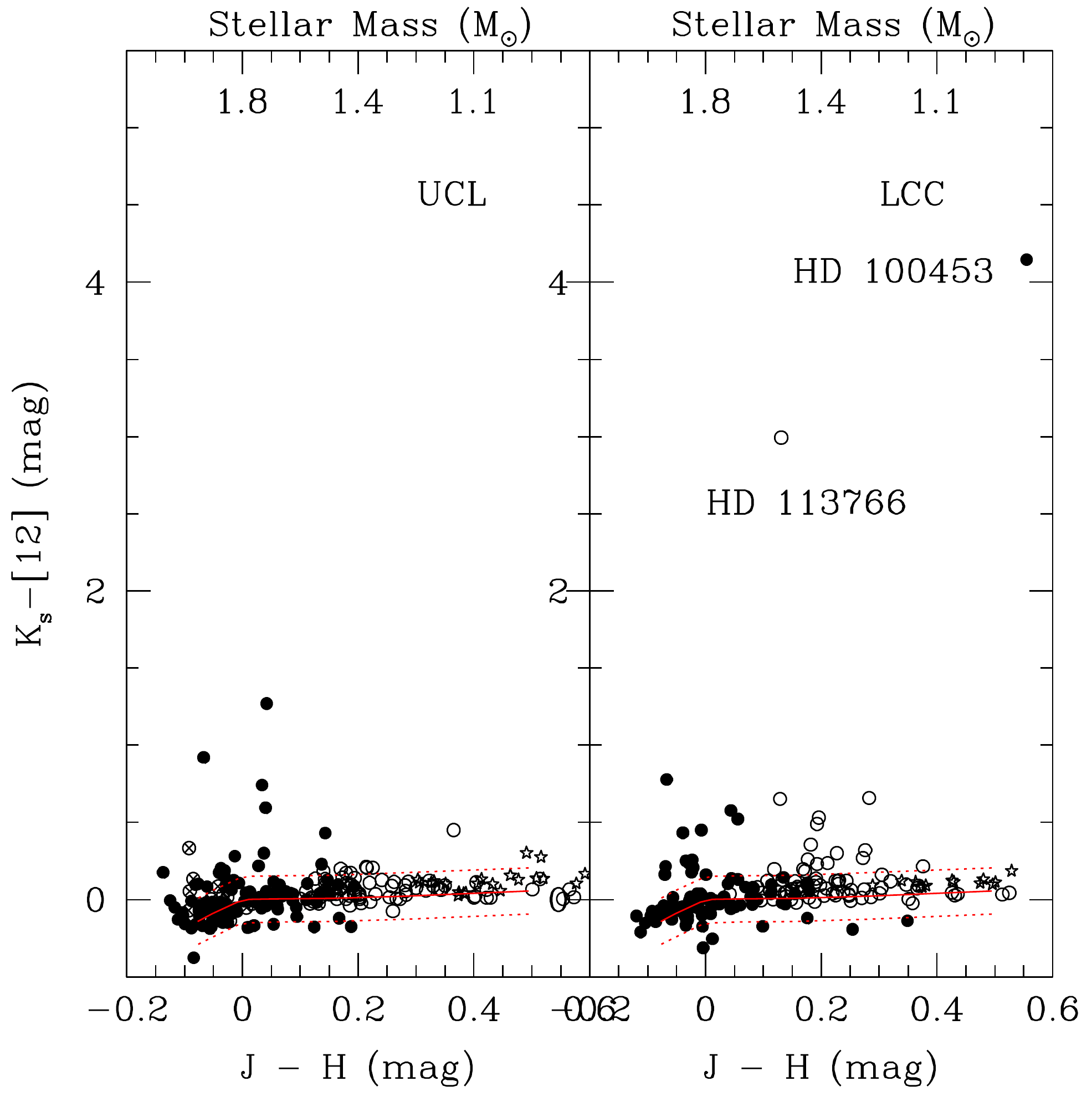}
\figcaption{K$_{s}$ - [12] color plotted as a function of J - H color for subgroups of Sco Cen.  The symbols are as described in Figure~\ref{fig:k24}. The dashed lines show the 3$\sigma$ range in K$_{s}$ - [12] color (0.23 mag) around the main sequence.
\label{fig:k12}}
\end{figure}

\begin{figure}
\figurenum{16}
\plotone{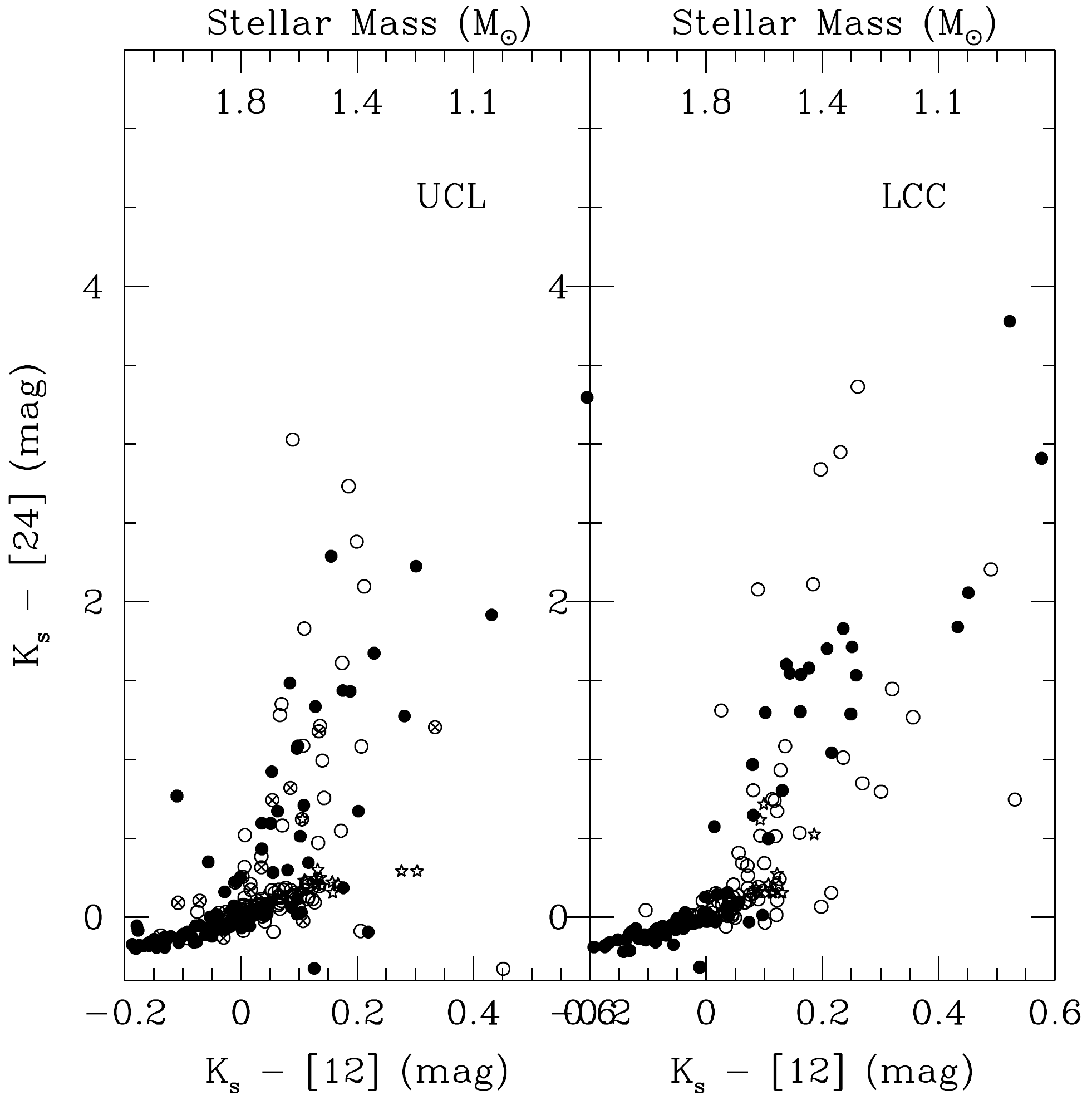}
\figcaption{K$_{s}$ - [12] color plotted as a function of $K_{s}$ - [24] color for subgroups of Sco Cen.  The symbols are as described in Figure~\ref{fig:k24}. 
\label{fig:k12k24}}
\end{figure}

\begin{figure}
\figurenum{17}
\plottwo{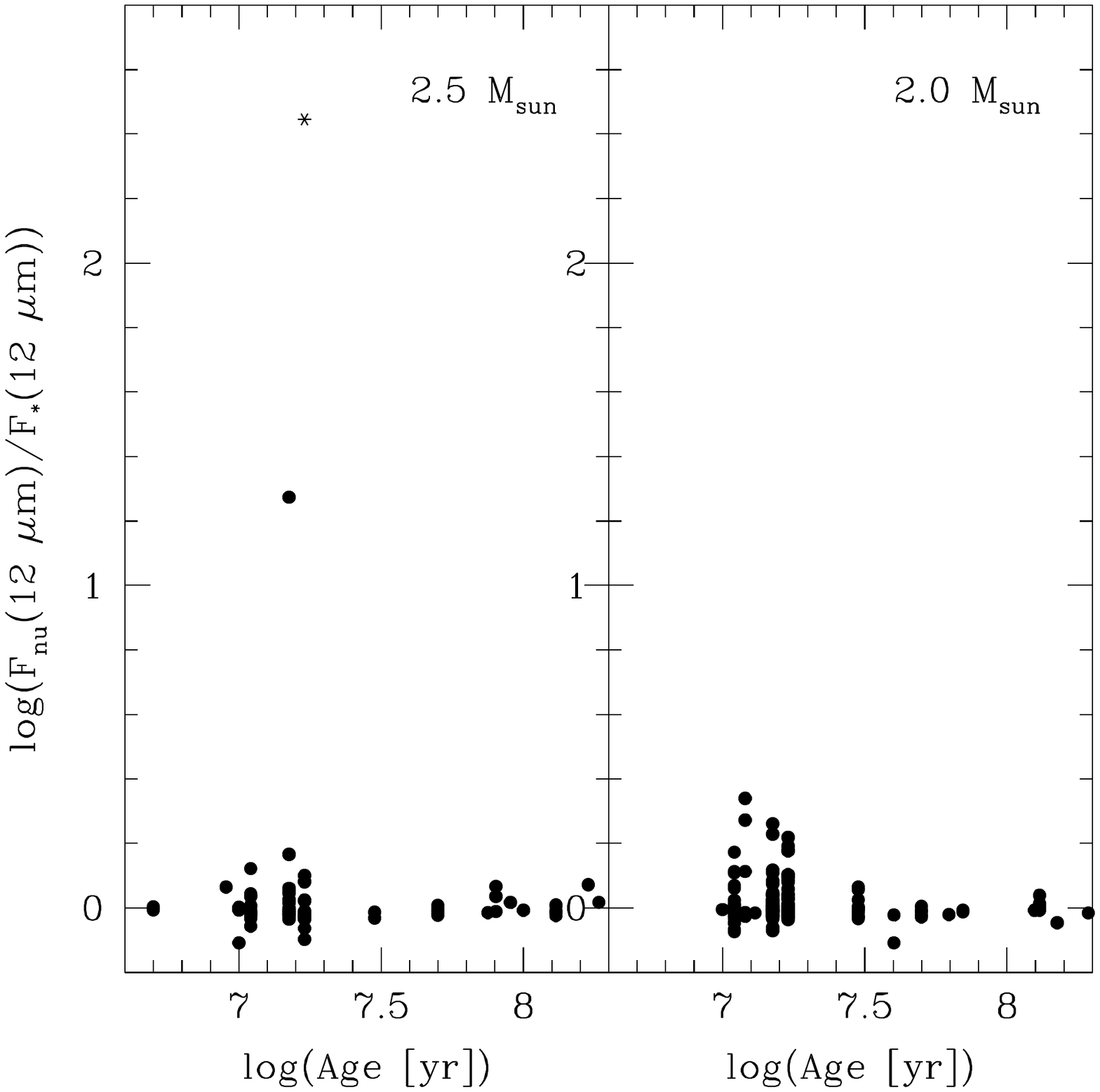}{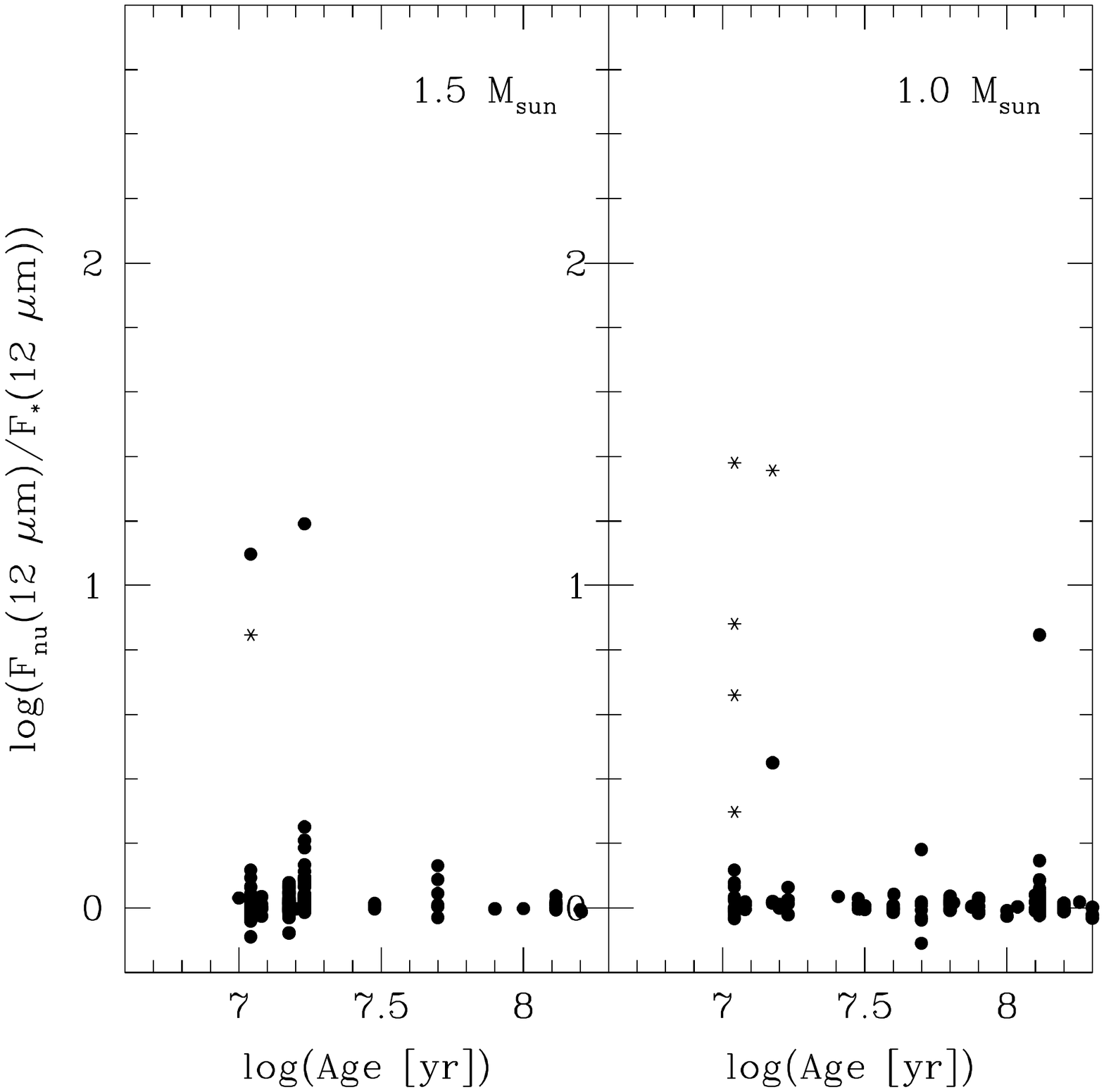}
\figcaption{The 12 $\mu$m flux ratio ($F_{\nu}$(12 $\mu$m)/$F_{*}$(12 $\mu$m)) as a function of stellar age. The sample of 2.0-2.5 $M_{\sun}$ stars is described in Table~\ref{tab:clusters}); the sample of 1.0-1.5 $M_{\sun}$ stars is described in \cite{chen11}. Objects with near-infrared (e.g., IRAC) and MIPS 24 $\mu$m excesses are plotted using asterisks; objects with only MIPS 24 $\mu$m excesses are plotted using solid circles. 
\label{fig:r12vage}}
\end{figure}

The WISE 12 $\mu$m band does provide a new opportunity to survey all of the ScoCen members at an additional wavelength. The 12 $\mu$m photometry can be used to probe warm terrestrial temperature dust ($\sim$240 K) at closer distances than the cooler dust detected at 24 $\mu$m ($\sim$120 K). The 12 $\mu$m excess associated with ScoCen members is significantly smaller than the 24 $\mu$m excess (see Figure \ref{fig:k12}), indicating that the inner regions of the disks possess less dust than the outer disk but that the terrestrial planet zone has not been completely cleared of parent bodies. We plotted $K_{s}$-[24] color as a function of $K_{s}$-[24] color (see Figure \ref{fig:k12k24}) to determine whether any trends exist in the infrared properties. We find that (1) stars without infrared excess appear to fall on a line that passes through the origin in $K_{s}$-[24] versus $K_{s}$-[12] space; (2) the majority of stars with $K_{s}$-[24] $>$0.5 excess typically also possess weak $K_{s}$-[12] $>$0.1 excess; and (3) UCL possesses more stars with weak 12 and 24 $\mu$m excess. Follow-up \emph{Spitzer} IRS spectra are required to characterize in detail the temperature distribution of the dust and therefore the spatial distribution of dust in these systems. We also plotted the time evolution of the normalized 12 $\mu$m fluxes, $F_{\nu}$(12 $\mu$m)/$F_{*}$(12 $\mu$m), (see Figure \ref{fig:r12vage}) for young ($<$300 Myr), nearby ($\lesssim$200 pc) stars using the same sample as described in Table \ref{tab:clusters}) for 2.0-2.5 M$_{\sun}$ stars and the same sample as described by \cite{chen11} for 1.0-1.5 M$\sun$ stars. Our analysis indicates that the production of warm terrestrial dust has subsided by an age of $\sim$30 Myr and $\sim$100 Myr for intermediate- and solar-mass stars, respectively, approximately the same timescale as observed for colder dust. The similarity in warm and cold dust dissipation timescales may indicate the presence of not only inner and outer disks of planetesimals but also giant planets at Jupiter- Saturn distances that are in stable or weakly unstable systems \citep{raymond11}.

\section{Conclusions}
We have obtained \emph{Spitzer} MIPS 24 and 70 $\mu$m photometry of 215 candidate B- and A-type members of Scorpius-Centaurus. We conclude the following:

1. The ScoCen subgroups US, UCL, and LCC possess statistically indistinguishable B+A 24 $\mu$m excess fractions of 25$^{+6}_{-5}$\%, 27$^{+4}_{-4}$\%, and 24$^{+5}_{-4}$\%, somewhat lower than the disk fractions observed by \cite{chen11} for F+G stars.

2. The $F_{\nu}$(24 $\mu$m)/$F_{*}$(24 $\mu$m) and $F_{\nu}$(70 $\mu$m)/$F_{*}$(70 $\mu$m) excesses of B+A stars is systematically smaller than that measured toward F+G stars, consistent with recent WISE observations \citep{rizzuto12}. Estimates of fractional infrared luminosity and grain temperature suggest that $L_{IR}/L_{*}$ decreases with increasing stellar mass and $T_{gr}$ is not dependent on stellar mass.

3. For 2.5 $M_{\sun}$ stars, the debris disk fraction does not appear to change statistically between 10 Myr and 15-20 Myr; however the 1st quartile of the MIPS $F_{\nu}$(24 $\mu$m)/$F_{*}$(24 $\mu$m) increases as expected from collisions between oligarchs at 30-150 AU in self-stirred disks. For 2.0 $M_{\sun}$ stars, the disk fraction does not appear to change statistically between 10 Myr and 15-20 Myr and the 1st quartile of the MIPS $F_{\nu}$(24 $\mu$m)/$F_{*}$(24 $\mu$m) increases only if primordial disks are retained in the sample.  

4. The known fraction of stellar companions at 10 - 200 AU around B+A stars is approximately a factor of two higher than that around F+G stars; therefore, the lower disk fraction and the lower fractional infrared excesses associated with detected disks may be the result of disk truncation.

5. ScoCen B- through G-type members with MIPS 24 $\mu$m excess also possess weak WISE 12 $\mu$m excess, indicating the presence of additional warm dust in the terrestrial planet forming zone, suggesting that terrestrial planets may still be forming around ScoCen stars at 10 - 100 Myr.

We would like to thank D. Hines, G. Kennedy, S. Kenyon, S. Lubow, and M. Wyatt for their helpful comments and suggestions. This work is based on observations made with the \emph{Spitzer Space Telescope}, which is operated by JPL/Caltech under a contract with NASA. Support for this work was provided by NASA through an award issued by JPL/Caltech. This research made use of the SIMBAD database, operated at CDS, Strasbourg, France, and data products from the 2MASS, which is a joint project of the U. Massachusetts and the Infrared Processing and Analysis Center/Caltech, funded by NASA and the NSF.

\bibliography{ms.bbl}

\begin{deluxetable}{llcccccc}
\tabletypesize{\scriptsize}
\tablecaption{Stellar Properties \label{tab:starprops}} 
\tablewidth{0pt}
\tablehead{
    \colhead{HIP} &
    \colhead{Name} &
    \colhead{Spectral} &
    \colhead{Distance} &
    \colhead{$A_{V}$} &
    \colhead{Subgroup} &
    \colhead{Separation} &
    \colhead{Notes} \\
    \omit &
    \omit &
    \colhead{Type} &
    \colhead{(pc)} &
    \colhead{(mag)} &
    \omit &
    \colhead{(arcsec)} &
    \omit \\
}
\tablewidth{0pt}
\tablecolumns{7}
\startdata
50083	&	HD 88981	&	A4mA6-F0	(8) & 	94.3	$_{	-1.5	}^{+	1.5	}$ &	0.149	$\pm$	0.047	&	LCC	\\
50520	&	HD 89715	&	A1V	(2) & 	80.3	$_{	-3.1	}^{+	3.4	}$ &	0.006	$\pm$	0.018	&	LCC	& 2.1 (a,b,c) & VB \\
50847	&	HD 90264	&	B8V	(8) & 	123.2	$_{	-2.7	}^{+	2.8	}$ &	0.026	$\pm$	0.017	&	LCC	& & SB \\
51991	&	HD 92315	&	A1Vp:Cr:Sr:Eu:	(8) & 	215.1	$_{	-23.5	}^{+	30.0	}$ &	0.180	$\pm$	0.012	&	LCC	\\
52357	&	HD 92896	&	A3IV	(8) & 	122.5	$_{	-9.9	}^{+	11.9	}$ &	0.349	$\pm$	0.019	&	LCC	& 0.53, 10.04 (a,b,c) & VB \\
53524	&	HD 95086	&	A8III	(8) & 	90.4	$_{	-3.2	}^{+	3.5	}$ &	0.000	$\pm$	0.019	&	LCC	& 4.87 (b) & VB \\
53701	&	HD 95324	&	B8IV	(8) & 	118.2	$_{	-7.3	}^{+	8.4	}$ &	0.044	$\pm$	0.017	&	LCC	& 3.88, 6.57 (a,b) & VB \\
54231	&	HD 96338	&	A0V	(9) & 	117.9	$_{	-6.3	}^{+	7.1	}$ &	0.031	$\pm$	0.054	&	LCC	\\
\enddata
\tablecomments{Upper Centaurus Lupus (UCL), Lower Centaurus Crux (LCC)}
\tablecomments{Common Proper Motion Binary (CPM), Eclipsing Binary (EB), Spectroscopic Binary (SB), Visual Binary (VB)}
\tablerefs{ (1) \cite{buscombe65}; (2) \cite{corbally84}; (3) \cite{gahm83}; (4) \cite{garrison67}; (5) \cite{garrison94}; (6) \cite{gray87}; (7) \cite{gray89}; (8) \cite{houk75}; (9) \cite{houk78}; (10) \cite{houk82}; (11) \cite{hiltner69}; (12) \cite{hube70}; (13) \cite{jaschek68}; (14) \cite{jordi97}; (15) \cite{levato96}; (16) \cite{morgan78}; (17) \cite{morris61}; (18) \cite{osawa65}; (19) \cite{paunzen01}; (20) Pecaut 2012, private communication; (21) \cite{schild71}; (22) \cite{upgren72}; (23) \cite{vieira03}; (24) \cite{wood77}}
\tablerefs{(a) \cite{dommanget02}; (b) \cite{kouwenhoven05}; (c) \cite{kouwenhoven07}; (d) \cite{levato87}; (e) \cite{shatsky02}}
\end{deluxetable}

%%%%%%%%%%%%%%%%%%%%%%%%%%%%%%%%%%%%%%%%%%%%%%%%%%%%%%%%%%%%%%%%%%%%%%

\clearpage
\begin{deluxetable}{llcccccccccccc}
\tabletypesize{\scriptsize}
\tablecaption{MIPS 24 $\mu$m and 70 $\mu$m Fluxes (Not Color-Corrected) \label{tab:mips}} 
\tablewidth{0pt}
\tablehead{
    \omit &
    \omit &
    \omit &
    \colhead{Measured} &
    \colhead{Measured} &
    \colhead{Predicted} &
    \omit &
    \colhead{Measured} &
    \colhead{Measured} &
    \colhead{Predicted} &
    \omit \\
    \colhead{HIP} &
    \colhead{Name} &
    \colhead{AOR} & 
    \colhead{F$_{\nu}$(24 $\mu$m)} &
    \colhead{$\sigma_{F24}^{\ddagger}$} &
    \colhead{F$_{\nu}$(24 $\mu$m)} &
     \colhead{$\chi_{24}$} &
   \colhead{F$_{\nu}$(70 $\mu$m)} &
    \colhead{$\sigma_{F70}^{\ddagger}$} &
    \colhead{F$_{\nu}$(70 $\mu$m)} &
    \colhead{$\chi_{70}$} \\
    \omit &
    \colhead{ID} &
    \colhead{(mJy)} &
    \colhead{(mJy)} &
    \colhead{(mJy)} &
    \omit &
    \colhead{(mJy)} &
    \colhead{(mJy)} &
    \colhead{(mJy)} &
    \omit \\
}
\tablewidth{0pt}
\tablecolumns{13}
\startdata
50083	&	HD088981	&	22767616	&	99.5	&	0.1	&	100.1	&	-0.2	&	$<$23.9	&	\nodata	&	10.9	&	\nodata	\\
50520$^{\dagger}$	&	HD089715	&	22767872	&	37.0	&	0.1	&	42.7	&	-3.8	&	$<$16.8	&	\nodata	&	4.6	&	\nodata	\\
50847	&	HD090264	&	22768128	&	44.2	&	0.2	&	47.9	&	-2.2	&	$<$22.6	&	\nodata	&	5.1	&	\nodata	\\
51991	&	HD092315	&	22768384	&	4.5	&	0.1	&	4.5	&	0.1	&	$<$14.6	&	\nodata	&	0.5	&	\nodata	\\
52357$^{\dagger}$	&	HD092896	&	22768640	&	14.2	&	0.2	&	14.6	&	-0.7	&	$<$13.2	&	\nodata	&	1.6	&	\nodata	\\
53524$^{\dagger}$	&	HD95086	&	16106496	&	45.6	&	0.1	&	13.7	&	31.1	&	654.6	&	5.4	&	1.5	&	14.7	\\
53701$^{\dagger}$	&	HD095324	&	22768896	&	16.7	&	12.7	&	20.8	&	-0.3	&	$<$826.3	&	\nodata	&	2.2	&	\nodata	\\
54231	&	HD096338	&	22769152	&	14.4	&	0.1	&	13.8	&	1.3	&	$<$5.5	&	\nodata	&	1.5	&	\nodata	\\
\enddata
\tablenotetext{\dag}{Source suffers some contamination at 24 $\mu$m.}
\tablenotetext{\ddag}{$\sigma_{F24}$ and $\sigma_{F70}$ are the statistical uncertainties in the 24 and 70 $\mu$m photometry. The total uncertainty can be calculated by adding the systematic, repeatability, and calibration uncertainties in quadrature (see Section 2.1).}
\end{deluxetable}

%%%%%%%%%%%%%%%%%%%%%%%%%%%%%%%%%%%%%%%%%%%%%%%%%%%%%%%%%%%%%%%%%%%%%%

\clearpage
\begin{deluxetable}{lccccccccccc}
\tabletypesize{\scriptsize}
\tablecaption{Single Temperature Black Body Model Parameters \label{tab:BBmodel}} 
\tablewidth{0pt}
\tablehead{
    \colhead{HIP} &
    \colhead{Name} &
    \colhead{$T_{eff}$} & 
    \colhead{$L_{*}$} &
    \colhead{$M_{*}$} &
    \colhead{$t_{age}$} &
    \colhead{$T_{gr}$} &
    \colhead{$L_{IR}/L_{*}$} &
    \colhead{$a_{min}$} &
    \colhead{$D$} &
    \colhead{$M_{dust}$} &
    \colhead{$M_{PB}$} \\
    \omit &
    \omit & 
    \colhead{(K)} &
    \colhead{($L_{\sun}$)} &
    \colhead{($M_{\sun}$)} &
    \colhead{(Myr)} &
    \colhead{(K)} &
    \omit &
    \colhead{($\mu$m)} &
    \colhead{(AU)} &
    \colhead{($M_{moon}$)} &
    \colhead{($M_{moon}$)} \\
}
\tablewidth{0pt}
\tablecolumns{10}
\startdata
\cutinhead{Lower Centaurus Crux}
53524	&	HD 95086	&	7500	&	7.2$^{+0.6}_{-0.5}$	&	1.6	&	17	&	63.7	&	1.4$\times$10$^{-3}$	&	0.8	&	270	&	1.5	&	$>$1 \\
55188	&	HD 98363	&	8770	&	11.4$^{+1.7}_{-1.5}$	&	1.9	&	17	&	132	&	7.1$\times$10$^{-4}$	&	2.6	&	17.2	&	0.011	&	$>$1	\\
56354	&	HD 100453	&	7500	&	10.5$^{+1.9}_{-1.6}$	&	1.6	&	17	&	40	&	3.9$\times$10$^{-2}$	&	1.4	&	829	&	700	& $>$52	\\
%%56379	&	HD 100546	& 	10500&	27.6$^{+2.5}_{-2.2}$	&	2.4	&	17	&	\\
57809	&	HD 102981	&	9550	&	38.2$^{+7.4}_{-6.2}$	&	2.1	&	17	&	78.2	&	8.4$\times$10$^{-5}$	&	7.9	&	69.7	&	0.066	&	$>$0.4	\\
58720	&	HD 104600	&	11614	&	68.7$^{+7.7}_{-6.9}$	&	2.7	&	17	&	126	&	7.7$\times$10$^{-5}$	&	8.7	&	32.6	&	0.0015	&	$>$0.7	\\
59282	&	HD 105613	&	8551	&	11.6$^{+1.6}_{-1.4}$	&	1.8	&	17	&	$>$111	&	4.4$\times$10$^{-5}$	&	2.6	&	$<$27.0	&	0.26	&	$>$0.06	\\
59397	&	HD 105857	&	8770	&	14.0$^{+2.5}_{-2.1}$	&	1.9	&	17	&	$>$167	&	9.0$\times$10$^{-5}$	&	2.8	&	$<$10.6	&	0.33	&	$>$0.2	\\
59502	&	HD 106036	&	8770	&	13.8$^{+1.4}_{-1.3}$	&	1.9	&	17	&	$>$109	&	2.2$\times$10$^{-4}$	&	2.8	&	$<$31.0	&	0.25	&	$>$0.4	\\
\enddata
\end{deluxetable}

%%%%%%%%%%%%%%%%%%%%%%%%%%%%%%%%%%%%%%%%%%%%%%%%%%%%%%%%%%%%%%%%%%%%%%

\clearpage
\begin{deluxetable}{lccc|cc|cc|c}
\tabletypesize{\scriptsize}
\tablecaption{Spitzer Intermediate-Age Disk Surveys \label{tab:clusters}} 
\tablewidth{0pt}
\tablehead{
    \omit &
    \omit &
    \colhead{Avg} &
    \omit &
    \multicolumn{2}{c}{2.0 $M_{\sun}$} &
    \multicolumn{2}{c}{2.5 $M_{\sun}$} &
    \omit \\
    \colhead{Region} & 
    \colhead{Age} &
    \colhead{Distance} &
    \colhead{Selection} &
    \colhead{$N_{stars}$} &
    \colhead{Range} &
    \colhead{$N_{stars}$} &
    \colhead{Range} &
    \colhead{Photometry} \\
    \omit & 
    \colhead{(Myr)} &
    \colhead{(pc)} &
    \colhead{Criteria} &
    \omit &
    \omit &
    \omit &
    \omit &
    \colhead{Reference} \\
}
\tablewidth{0pt}
\tablecolumns{7}
\startdata
Upper Sco & 10 &$\sim$146& SpT & 24 & A0-A9 & 16 & B9-B9.5 & 1 \\
$\beta$ Pic MG & 12 &$\lesssim$60& SpT & 5 & A0-A7 & 1 & B9-B9.5 & 4 \\
UCL & 15 &$\sim$142 & SpT & 6 & A0-A7 & 3 & B9-B9.5 & 6 \\
& & & SpT & 44 & A0-A7 & 16 & B9-B9.5 & 8 \\
LCC & 17 &$\sim$118& SpT & 46 & A0-A7 & 15 & B9-B9.5 & 8 \\
Tuc-Hor & 30 & $\lesssim$30 & SpT & 12 & A0-A7 & 2 & B9-B9.5 & 7 \\
Argus & 40 & $\lesssim$150 & SpT & 5 & A0-A7 & \nodata & B9-B9.5 & 7 \\
IC 2391 & 50 &$\sim$150& SpT & 5 & A0-A7 & \nodata & B9-B9.5 & 5 \\
\enddata
\tablerefs{(1) \citet{carpenter09}; (2) \citet{gorlova06}; (3) \citet{plavchan09}; (4) \citet{rebull08}; (5) \citet{siegler07}; (6) \citet{su06}; (7) \citet{zuckerman11}; (8) this work}  
%(3) \citet{flaherty08}; (4) \citet{hernandez06}; (5) \citet{hernandez07}; (6) \citet{hernandez08}; (7) \citet{lada06}; (9) \citet{rieke05}; 
\tablenotetext{a}{Photosphere estimates made based on \emph{Spitzer} IRAC photometry published in \citet{stauffer05}.}
\end{deluxetable}

%%%%%%%%%%%%%%%%%%%%%%%%%%%%%%%%%%%%%%%%%%%%%%%%%%%%%%%%%%%%%%%%%%%%%%

\begin{deluxetable}{lcccc}
\tabletypesize{\scriptsize}
\tablecaption{Observations of 2 - 2.5 $M_{\sun}$ Stars in US \label{tab:us}} 
\tablewidth{0pt}
\tablehead{
   \colhead{HIP} & 
   \colhead{SpT} &
   \colhead{$F_{\nu}$(24 $\mu$m)/$F_{*}$(24 $\mu$m)} &
   \colhead{Type} &
   \colhead{Notes} \\
}
\tablewidth{0pt}
\tablecolumns{5}
\startdata
\cutinhead{2.5 M$_{\sun}$ Stars}
76701 & B9V & 0.98$\pm$0.07 & No Excess (1) \\
76633 & B9V & 1.30$\pm$0.07 & No Excess (1) \\
78530 & B9V & 1.07$\pm$0.07 & No Excess (1) \\
78549 & B9.5V & 1.03$\pm$0.07 & No Excess (1) \\
78809 & B9V & 1.02$\pm$0.07 & No Excess (1) \\
78956 & B9.5V & 1.55$\pm$0.07 & Debris (1) \\
78968 & B9V & 1.84$\pm$0.07 & Debris (1) \\
79124 & A0V & 1.06$\pm$0.07 & No Excess (1) \\
\enddata
\tablerefs{(1) \cite{carpenter09}; (2) \cite{chen11}; (3) this work; (4) \cite{oudmaijer92}; (5) \cite{vieira03}}
\end{deluxetable}

%%%%%%%%%%%%%%%%%%%%%%%%%%%%%%%%%%%%%%%%%%%%%%%%%%%%%%%%%%%%%%%%%%%%%%

\begin{deluxetable}{lcccc}
\tabletypesize{\scriptsize}
\tablecaption{Observations of 2 - 2.5 $M_{\sun}$ Stars in UCL \label{tab:ucl}} 
\tablewidth{0pt}
\tablehead{
   \colhead{HIP} & 
   \colhead{SpT} &
   \colhead{$F_{\nu}$(24 $\mu$m)/$F_{*}$(24 $\mu$m)} &
   \colhead{Type} &
   \colhead{Notes} \\
}
\tablewidth{0pt}
\tablecolumns{5}
\startdata
\cutinhead{2.5 M$_{\sun}$ Stars}
69113 & B9V & 0.96$\pm$0.05 & No Excess (2) \\ 
69749 & B9IV & 1.12$\pm$0.04 & No Excess (3) \\
69845 & B9V & 1.03$\pm$0.04 & No Excess (3) \\
70690 & B9V & 0.96$\pm$0.04 & No Excess (3) \\
71724 & B8/9V & 0.95$\pm$0.04 & No Excess (3) \\
73266 & B9V & 0.90$\pm$0.04 & No Excess (3) \\
75151 & B9IVSi(SrCr) & 1.01$\pm$0.17 & No Excess (2) \\
75210 & B8/9V & 2.97$\pm$0.07 & Debris (2) \\
\enddata
\tablerefs{(1) Chen et al. (2011); (2) Su et al. (2006); (3) this work}
\end{deluxetable}

%%%%%%%%%%%%%%%%%%%%%%%%%%%%%%%%%%%%%%%%%%%%%%%%%%%%%%%%%%%%%%%%%%%%%%

\begin{deluxetable}{lcccc}
\tabletypesize{\scriptsize}
\tablecaption{Observations of 2 - 2.5 $M_{\sun}$ Stars in LCC \label{tab:lcc}} 
\tablewidth{0pt}
\tablehead{
   \colhead{HIP} & 
   \colhead{SpT} &
   \colhead{$F_{\nu}$(24 $\mu$m)/$F_{*}$(24 $\mu$m)} &
   \colhead{Type} &
   \colhead{Notes} \\
}
\tablewidth{0pt}
\tablecolumns{5}
\startdata
\cutinhead{2.5 M$_{\sun}$ Stars}
56379 & B9Vne & 5105$\pm$150 &  Protoplanetary (2) & Herbig Be (4) \\
58452 & B8/9V & 1.01$\pm$0.04 & No Excess (3) \\
58720 & B9V & 4.40$\pm$0.04 & Debris (3) \\
60183 & B9V & 4.47$\pm$0.04 & Debris (3) \\
61796 & B8V & 0.97$\pm$0.04 & No Excess (3) \\
62058 & B9Vn & 0.96$\pm$0.04 & No Excess (3) \\
65021 & B9V & 0.98$\pm$0.04 & No Excess (3) \\
65178 & B9V & 0.93$\pm$0.04 & No Excess (3) \\
\enddata
\tablerefs{(1) \cite{chen11}; (2) \cite{oudmaijer92}; (3) this work; (4) \cite{vieira03}}
\end{deluxetable}

%%%%%%%%%%%%%%%%%%%%%%%%%%%%%%%%%%%%%%%%%%%%%%%%%%%%%%%%%%%%%%%%%%%%%%

\begin{deluxetable}{lcccccccc}
\tabletypesize{\scriptsize}
\tablecaption{$F_{\nu}$(24 $\mu$m)/$F_{*}$(24 $\mu$m) Statistics for ScoCen Evolutionary Sample \label{tab:statistics}}
\tablewidth{0pt}
\tablehead{
        \colhead{Group} & 
        \colhead{Age}    &
        \colhead{Mean} &
        \colhead{Std Dev}    & 
        \colhead{3rd Q} &
        \colhead{Median}    & 
        \colhead{1st Q}    & 
        \colhead{RS Z} &
        \colhead{RS Prob} \\
        \omit &
        \colhead{(Myr)}   & 
        \omit &         
        \omit & 
        \omit &         
        \omit & 
        \omit & 
        \omit \\
}
\tablewidth{0pt}
\tablecolumns{9}
\startdata
\cutinhead{2.5 M$_{\sun}$ Stars - All}
US  & 10 & 30 & 100 & 1.03 & 1.11 & 1.99 & 0 (-1.6) & 1 (0.06) \\
UCL & 15 & 1.7 & 1.4 & 0.95 & 1.03 & 1.97 & 1.6 (0) & 0.06 (1) \\
LCC & 17 & 500 & 1600 & 0.97 & 1.01 & 4.47 & 0.8 (0.9) & 0.2 (0.2) \\
\cutinhead{2.5 M$_{\sun}$ Stars - Primordial Removed}
US  & 10 & 1.5 & 0.9 & 1.03 & 1.07 & 1.84 &  0 (-1.3) & 1 (0.1) \\
UCL & 15 & 1.7 & 1.4 & 0.95 & 1.03 & 1.97 & 1.3 (0) & 0.1 (1) \\
LCC & 17 & 2.2 & 1.9 & 0.97 & 0.98 & 4.47 & 1.1 (0.5) & 0.1 (0.3) \\
\cutinhead{2.0 M$_{\sun}$ Stars - All}
US  & 10 & 22 & 86 & 1.00 & 1.04 & 3.36 & 0 (0.9) & 1 (0.2) \\
UCL & 15 & 1.8 & 2.9 & 0.97 & 1.06 & 1.31 & -0.9 (0) & 0.2 (1) \\
\enddata
\tablecomments{Statistics comparing the populations of US, UCL, and LCC. We calculate the mean  F$_{\nu}$(24 $\mu$m)/F$_{*}$(24 $\mu$m) for each subgroup (including and excluding "primordial disks"), the standard deviation ($\sigma$), the 1st quartile, the median, the 3rd quartile, and the Wilcoxan rank sum probability and Z parameter. For the Wilcoxon rank sum test parameters, the first (second) entry in the rank sum test statistics compares each population to US (UCL). A positive Z parameter means that the sample has a larger median value than US (first entry) and UCL (second entry). A low rank sum probability means that the median F$_{\nu}$(24 $\mu$m)/F$_{*}$(24 $\mu$m) of the populations are very different.}
\end{deluxetable}

%%%%%%%%%%%%%%%%%%%%%%%%%%%%%%%%%%%%%%%%%%%%%%%%%%%%%%%%%%%%%%%%%%%%%%

\end{document}